\PassOptionsToPackage{table}{xcolor}
\documentclass[11pt, a4paper]{sakana_format}

\usepackage{latexsym}
\usepackage[T1]{fontenc}
\usepackage[utf8]{inputenc}
\usepackage{graphicx}

\usepackage{hyperref}       %
\usepackage{url}            %
\usepackage{booktabs}       %
\usepackage{capt-of}        %
\usepackage{amsfonts}       %
\usepackage{nicefrac}       %
\usepackage{xcolor}         %
\definecolor{abccol}{rgb}{0.80,0.89,0.99}  %
\definecolor{vscol}{rgb}{0.96,0.96,0.93}   %

\usepackage{amsmath}
\usepackage{amssymb}
\usepackage{mathtools}
\usepackage{algorithm}
\usepackage{algorithmic}
\usepackage{multicol}  %
\usepackage{amsthm}
\usepackage{listings}

\lstdefinestyle{prompt}{
  basicstyle=\ttfamily\footnotesize,
  columns=fullflexible,
  breaklines=true,
  breakatwhitespace=true,
  frame=single,
  rulecolor=\color{black!25},
  backgroundcolor=\color{black!2},
  showstringspaces=false
}

\newcommand{\authornoteoff}[1]{}

\usepackage{subfigure}
\usepackage{needspace}       %

\newcommand{\papercodelink}{%
  \par\smallskip
  \begin{center}
    \small
    \raisebox{-0.15em}{\includegraphics[height=1.1em]{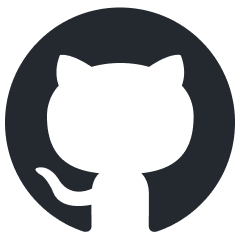}}%
    \hspace{0.45em}%
    \textbf{Code:} %
    \href{https://github.com/SakanaAI/QD-Rec}{\nolinkurl{github.com/SakanaAI/QD-Rec}}%
  \end{center}
  \smallskip
}

\title{QD-Rec: Open-Ended Evolution of Recommendations via Quality--Diversity}

\author[1]{Takashi Ishida}
\author[1]{Taro Makino}
\author[1]{Issa Sugiura}

\affil[1]{Sakana AI}

\begin{document}

\begin{abstract}
Open-ended search combined with foundation models has produced advances such as faster algorithms, new results on open mathematical problems, and self-improving agents. We extend this paradigm to recommendation for decision-making, focusing on decisions where the set of possible choices is fixed but the reasons for choosing them are not. In high-stakes decisions, e.g., which company to acquire or which customer to target, comparing diverse, well-reasoned options can be more useful than a single ranked list. We formalize \emph{open-ended recommendation} as building a diverse, high-quality set of recommendations, each an item together with a \emph{hypothesis}, a natural-language reason why the item fits the needs of a decision maker. We propose QD-Rec, a quality--diversity method that selects items with a contextual bandit and generates hypotheses with a foundation model from earlier ones kept in an archive as stepping stones. A new recommendation enters the archive if it is novel or better than its nearest neighbor. Across M\&A, customer targeting, and paper and book recommendation, QD-Rec gives the highest mean quality and builds the largest archive in most settings.
\end{abstract}
\maketitle

\papercodelink

\section{Introduction}

Open-ended search continually produces novel and increasingly capable or diverse discoveries, and each can serve as a stepping stone for further discovery \citep{stanley2015book,hughes2024icml}. Foundation models can act as the generator in this search, drawing on earlier candidates to write new ones. This approach has found candidates that no single query to the model would produce, including creative text \citep{bradley2023quality}, adversarial prompts for red teaming \citep{samvelyan2024rainbow}, tasks that reveal a model's capabilities \citep{lu2025automated}, loss functions \citep{lu2024discovering}, faster algorithms \citep{novikov2025alphaevolve}, new mathematical constructions \citep{romera2024funsearch,novikov2025alphaevolve}, agent designs \citep{hu2025iclr}, and self-improving agents \citep{zhang2026iclr,wang2026iclr,zhang2026arxiv}.

We ask whether open-ended search can extend to recommendation for decision-making. Unlike open-ended generation, where the model generates new candidates, many decisions are choices from a fixed pool of candidates: an acquirer chooses among the companies that exist, a reader among the papers already written, a marketer among the customers. At first glance, this leaves no room for open-endedness, since the pool is finite and known in advance. However, only the items are fixed. The same item can be chosen for many different reasons, so the search over reasons is still open-ended.

We therefore treat recommendation for such decisions as an open-ended search over item--hypothesis pairs. A \emph{hypothesis} is a natural-language explanation of why an item fits the decision maker's profile, such as the profile of one acquirer or one reader. The search returns a diverse, high-quality set of such pairs for a single profile, rather than one ranked list of items. We call this problem \emph{open-ended recommendation}. In high-stakes decisions, people need to compare candidate items together with the reasons for choosing them \citep{aggarwal2016recommender}. We evaluate the set by its quality and diversity, following the quality--diversity (QD) view of \emph{illuminating} a space of alternatives \citep{mouret2015illuminating}.

Open-ended recommendation has two properties that make open-ended search a good fit. First, evaluation is costly: we learn a recommendation's value only by writing its hypothesis and judging it, so we cannot simply score every possible recommendation in advance and sort them. The system must therefore decide which item to evaluate next and learn from each result. Second, the space is larger than the pool: each item allows unlimited hypotheses, so even evaluating every item once would leave most of the space unexplored. Good hypotheses often come from adapting or combining earlier ones, so keeping modest intermediate hypotheses as stepping stones can help \citep{lehman2011abandoning,stanley2015book}. QD and novelty search are well suited to such problems \citep{risi-et-al:book2025}. \citet{gallotta2024tg} explored QD for generating and recommending spaceship designs through grammar-based evolution guided by user preferences. We extend this connection between QD and recommendation to decisions involving a fixed pool of existing items, using foundation models to generate and refine hypotheses for recommending them.\footnote{Evolutionary methods have been applied to decision-making more broadly, for example to evolve neural networks that prescribe interventions against COVID-19 \citep[Section~6.4]{risi-et-al:book2025}.}

\begin{figure}[t]
\centering
\includegraphics[width=\linewidth]{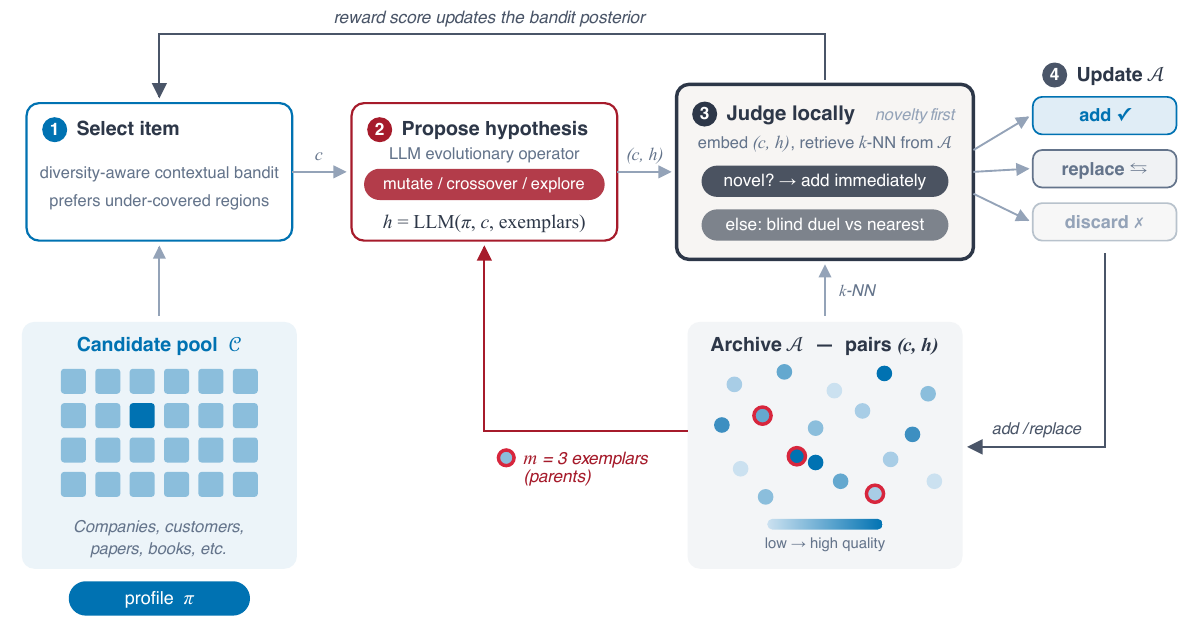}
\vspace{-4pt}
\caption{Overview of QD-Rec. Each iteration selects an item from the fixed pool $\mathcal{C}$ via a diversity-aware bandit, generates a hypothesis with an LLM conditioned on archive exemplars, which act as parents (an evolutionary operator), and judges the new pair against its $k$ nearest neighbors: a novel pair is added immediately; a non-novel one survives only by winning a pairwise duel against its neighbor.}
\label{fig:overview}
\vspace{-6pt}
\end{figure}

We introduce \textbf{QD-Rec}, a quality--diversity search method for open-ended recommendation (Figure~\ref{fig:overview}). In QD-Rec, an LLM called the \emph{proposer} writes hypotheses, and an LLM judge or humans score them. Each iteration has three steps. A contextual bandit selects an item from the candidate pool, balancing items that look promising against regions the archive has not yet covered. The proposer generates a hypothesis for that item, building on a few earlier hypotheses from the archive. The new recommendation is kept if it is novel, or if it is better than its nearest neighbor in the archive. Novelty and quality are thus judged \emph{locally}, by comparison with nearby recommendations.

Our contributions are: (i)~we formalize \emph{open-ended recommendation} as a quality--diversity search over item--hypothesis pairs, where the items come from a fixed pool; (ii)~we propose QD-Rec, with two ablations isolating its components (Table~\ref{tab:method-variants}); and (iii)~across four domains and two proposers, QD-Rec gives the highest mean quality in most LLM-judged settings, usually builds the largest archive, and outperforms diversity-aware rerankers on human-rated quality.

\section{Background and Related Work}
We briefly review quality--diversity methods and novelty search, evolutionary loops built around foundation models, and recommender systems.
\paragraph{Quality--diversity and novelty search.}
Quality--diversity (QD) methods aim to discover a \emph{set} of high-performing yet diverse solutions rather than a single optimum.
Multi-dimensional Archive of Phenotypic (MAP) Elites ``illuminates'' a search space by maintaining an archive and retaining elites across behavioral niches \citep{mouret2015illuminating}.
Novelty search emphasizes exploration by rewarding behavioral novelty, providing a mechanism for stepping-stone trajectories in which intermediate solutions enable later progress \citep{lehman2011abandoning}.
Recent work continues to revisit local competition and novelty definitions in QD, including dominated novelty variants \citep{bahlous2025dominated}, and to reduce dependence on hand-designed behavioral descriptors through methods such as AURORA \citep{grillotti2022unsupervised} and dimensionless MAP-Elites \citep{dharna2025foundation}.

\paragraph{Foundation models in evolutionary loops.}
A growing family of systems places a frozen large model inside an evolutionary or QD loop as a proposal operator or mutator, keeping an archive of artifacts that the model conditions on to propose the next: diverse creative text in QDAIF \citep{bradley2023quality}, adversarial prompts in Rainbow Teaming \citep{samvelyan2024rainbow}, code and program search in ELM \citep{lehman2023evolution} and FunSearch \citep{romera2024funsearch}, algorithm discovery in AlphaEvolve \citep{novikov2025alphaevolve}, sample-efficient program evolution in ShinkaEvolve \citep{lange2026iclr}, open-ended evaluation tasks in Automated Capability Discovery \citep{lu2025automated}, and agent designs in ADAS \citep{hu2025iclr}.\footnote{In evolutionary loops, foundation models can also play roles other than proposing new artifacts. In CycleQD \citep{kuroki2024agent} and AC/DC \citep{dai2026iclr}, they are themselves the artifacts being evolved. In a replication of Picbreeder \citep{earle2026gecco}, vision--language models take the place of its human evaluators.}
A separate line of work is \emph{self-referential}: STOP \citep{zelikman2023stop}, the Darwin G\"odel Machine \citep{zhang2026iclr}, the Huxley--G\"odel Machine \citep{wang2026iclr}, and Hyperagents \citep{zhang2026arxiv} turn the search on the machinery itself, so the improved scaffold or agent performs the next round of improvement. QD-Rec belongs to the first group, with item--hypothesis pairs as the artifacts. However, these systems create new artifacts, while QD-Rec must choose its items from a fixed pool.

\paragraph{Recommender systems.}
Recommenders learn either \emph{offline}, from a fixed dataset of past interactions \citep{aggarwal2016recommender}, or \emph{online}, as a sequential decision problem with an exploration--exploitation trade-off, usually modeled as a (contextual) bandit \citep{lattimore2020bandit,agarwal2009icdm,li2010www,agrawal2013icml,song2022wsdm}. Previous work has also combined bandits with QD to select parents or bins from an archive \citep{sfikas2021gecco,gallotta2024tg} and to choose how to generate new solutions \citep{cully2021gecco,gaier2020gecco}. Whereas these methods generate the candidate solutions themselves, QD-Rec uses a contextual bandit to select existing items rather than parents or generation strategies. QD-Rec then generates hypotheses for the selected items. As in QD, it is evaluated by its final archive.

Orthogonally, a long line of work treats diversity, novelty, coverage, and serendipity as first-class objectives \citep{mcnee2006being, ziegler2005improving, vargas2011rank, kaminskas2016diversity}, pursued offline by re-ranking \citep{carbonell1998use,chen2018dpp} and online by diversity-aware bandits \citep{radlinski2008icml,liu2020aaai}. These methods diversify the list shown to a user; QD-Rec instead enforces diversity at archive admission, following recent QD systems \citep{lu2025automated,dharna2025foundation}: an LLM judges whether each new pair is novel relative to its nearest archive neighbors.
Finally, explainable recommendation adds a natural-language explanation to each recommended item \citep{zhang2020explainable}, and so do LLM recommenders that reason before choosing or justify afterward. QD-Rec also pairs each item with a reason, but it does so inside an evolutionary loop.

\section{Open-Ended Recommendation}
\label{sec:problem}
Let $\mathcal{C}=\{c_1,\dots,c_N\}$ be a fixed candidate pool (e.g., companies for M\&A, customer profiles for targeting, papers, or books), and let $\pi$ be the \emph{profile} of one decision maker. A \emph{hypothesis} $h\in\mathcal{H}$ is a natural-language explanation of why an item fits $\pi$, and the hypothesis space $\mathcal{H}$ is unbounded. A \emph{recommendation} is a pair $(c,h)\in\mathcal{C}\times\mathcal{H}$, so the same item can be recommended several times with different hypotheses. Its quality $q(c,h)$ is a score from a human or an LLM judge, for example on a 0 to 10 scale. The score is known only after the hypothesis is written and evaluated, and each evaluation is costly. Given $\pi$ and $\mathcal{C}$, the task is to return a set $\mathcal{S}\subset\mathcal{C}\times\mathcal{H}$ of recommendations that have high quality and differ from one another, using at most $T$ evaluations. Section~\ref{sec:setup} defines the metrics for quality and diversity.

\section{QD-Rec}
\label{sec:method}
We present \textbf{QD-Rec}, an \emph{archive-based quality--diversity (QD) method} for open-ended recommendation. QD-Rec keeps an archive $\mathcal{A}$ of retained recommendations and returns it as $\mathcal{S}$ at the end of the run. Each new entry is compared only with its nearest neighbors in the archive to decide whether it is kept. Each archive entry $x$ stores a recommendation $(c,h)$, its embedding $e=\mathrm{Embed}(c,h)$, and its quality score $q(x)$. The embedding is computed from a structured item profile followed by the hypothesis text. We use cosine similarity $\mathrm{sim}(x_i,x_j)=\langle e_i,e_j\rangle/(\|e_i\|\,\|e_j\|)$ to define these neighborhoods.

\subsection{Local novelty, ranking, and archive update}
\label{sec:archive-update}
At each step we propose a new pair by selecting an item $c$ and generating a hypothesis $h$ (Section~\ref{sec:full-method}); we write $x_{\text{new}}=(c,h,e_{\text{new}})$ with $e_{\text{new}}=\mathrm{Embed}(c,h)$ for the proposed entry (its quality estimate is not yet known at this point). We retrieve the $k$ nearest neighbors from the archive,
\[
\mathcal{N}_k(x_{\text{new}})=\operatorname*{arg\,max}^{k}_{x\in\mathcal{A}}\,\mathrm{sim}(x_{\text{new}},x).
\]
The archive update then proceeds in the following two steps:
\begin{itemize}
\item \emph{Novelty check}: an LLM judge is prompted with the profile $\pi$, the new entry, and its $k$ nearest neighbors, and returns a binary decision $\texttt{is\_novel}\in\{0,1\}$ (with brief reasoning). A novel entry is admitted immediately; no comparison is made.
\item \emph{Local quality comparison} (only when not novel): we use a blind pairwise duel: of the $k$ neighbors only the single \emph{nearest} one is contested. It and $x_{\text{new}}$ are shuffled into two anonymous positions, pointwise scores are withheld, and the judge, never told which of the two is the newcomer, chooses the stronger under $\pi$ (ties are not allowed). If $x_{\text{new}}$ wins it replaces that neighbor; otherwise it is discarded. Algorithm~\ref{alg:qd-search} writes this comparison as $\mathrm{Better}_{\pi}$. Separating the two questions keeps the retention comparison \emph{blind} (the judge cannot favor or disfavor an entry for being new) and skips the comparison call entirely whenever the entry is novel.
\end{itemize}
This embedding-neighborhood admission scheme is inspired by recent novelty-search/QD systems that use learned representations for similarity and archive admission \citep{lu2025automated,dharna2025foundation}.

The quality score $q(x)$ of each evaluated pair (Section~\ref{sec:problem}) has two uses. The bandit reward $r(x)$ is the score divided by the maximum of its scale, so that rewards lie in $[0,1]$. Reported Quality averages the score over the final archive, on its original scale.

\subsection{Item selection and hypothesis generation}
\label{sec:full-method}
The archive mechanics above leave two decisions open: which item $c\in\mathcal{C}$ to evaluate next, and how to generate a hypothesis $h$ for it.

\paragraph{Item selection.}
Each candidate item $c$ has a \emph{precomputed candidate embedding} $\psi(c)\in\mathbb{R}^d$, distinct from the archive-entry embedding $e$ and taken to be L2-normalized. It embeds the item's text: the title and abstract of a paper, the title and description of a book, the business description of a company, or the profile of a customer. We model a scalar reward $r\in[0,1]$ by Bayesian linear regression, $r = \psi(c)^\top \theta + \epsilon$ with $\epsilon\sim\mathcal{N}(0,\sigma^2)$ and prior $\theta\sim\mathcal{N}(0,\lambda^{-1}I)$. For the set $\mathcal{D}$ of evaluated pairs we maintain the sufficient statistics
\[
A \;=\; \lambda I + \sigma^{-2}\!\!\sum_{(\psi,r)\in\mathcal{D}}\!\!\psi\psi^\top,
\qquad
b \;=\; \sigma^{-2}\!\!\sum_{(\psi,r)\in\mathcal{D}}\!\! r\,\psi,
\qquad
\theta\mid\mathcal{D}\;\sim\;\mathcal{N}(A^{-1}b,\; A^{-1}),
\]
and a new observation $(\psi(c), r)$ updates them in closed form:
\[
A \;\leftarrow\; A + \sigma^{-2}\,\psi(c)\psi(c)^\top,
\qquad
b \;\leftarrow\; b + \sigma^{-2}\, r\,\psi(c).
\]
Linear Thompson sampling \citep{agrawal2013icml} then draws $\tilde{\theta}\sim\mathcal{N}(A^{-1}b, A^{-1})$ at each step, and, with a small modification for diversity, we select the candidate that maximizes the \emph{diversity-aware selection score}
\begin{equation}
s(c) \;=\; \psi(c)^\top \tilde{\theta} \;+\; \beta\, d(c,\mathcal{R}),
\qquad
d(c,\mathcal{R}) \;=\; 1-\max_{c'\in\mathcal{R}} \psi(c)^\top \psi(c'),
\label{eq:selection-score}
\end{equation}
where $\mathcal{R}$ collects the candidate embeddings of the current archive. The bonus $d(c,\mathcal{R})$ is the cosine distance to the \emph{nearest} covered item (the standard nearest-neighbor novelty measure), so it is large exactly when nothing similar has been archived yet, and $\beta\ge 0$ trades this coverage of under-explored regions against the sampled quality estimate $\psi(c)^\top\tilde{\theta}$. The bandit \emph{reward} is unchanged: after evaluation we update the posterior using the normalized reward $r(x)$ defined above, so $\beta$ affects only \emph{which} items are chosen for evaluation, steering the archive to spread rather than cluster. Setting $\beta=0$ recovers the plain quality-only bandit; we use $\beta=1$ throughout. We warm-start the posterior from the seed archive: each seed entry contributes one observation $(\psi(c), r(x))$, using the same normalization.

An item can be selected more than once, because a new hypothesis for it can receive a different quality score. The same item can also appear in the archive several times.\footnote{For an item already in the archive, the bonus $d(c,\mathcal{R})$ is zero, so QD-Rec rarely selects it again. In our runs, no final QD-Rec archive contains the same item twice.} If a new hypothesis for an item differs enough from the entries near it, the novelty check admits it as a new entry.

\paragraph{Hypothesis generation (evolutionary operator).}
Given the selected item $c$, we prompt the proposer LLM with the profile $\pi$, the item $c$, and $m$ exemplars sampled uniformly from the current archive. The prompt presents the exemplars as existing hypotheses and instructs the model to apply one of three operators, \emph{mutate} (adapt one exemplar's angle to $c$), \emph{crossover} (combine angles from several exemplars), or \emph{explore} (propose an angle the exemplars miss), so the archive acts as an evolving population \citep{guo2024iclr,lehman2023evolution}.

\paragraph{Simpler variants.}
We also define two simpler variants of QD-Rec, which serve as ablations of the full method. \emph{Novelty+Evo} replaces the bandit with uniform-random item selection. \emph{Novelty} removes both the bandit and the exemplars, so the proposer generates hypotheses zero-shot. All three methods share the archive update of Section~\ref{sec:archive-update}, and Table~\ref{tab:method-variants} compares them. Algorithm~\ref{alg:qd-search} gives QD-Rec in full.

\begin{algorithm*}[!t]
\caption{QD-Rec. The Novelty+Evo ablation uses uniform-random selection from $\mathcal{C}$; Novelty additionally uses no exemplars ($m=0$).}
\label{alg:qd-search}
\small
\begin{algorithmic}[1]
\REQUIRE Candidate pool $\mathcal{C}$, profile $\pi$, seed archive $\mathcal{A}$, iterations $T$, neighborhood size $k$, exemplars $m$, diversity weight $\beta\ge0$, prior precision $\lambda$, noise variance $\sigma^2$; warm-start $A=\lambda I$, $b=0$ with one observation $(\psi(c),r(x))$ per seed entry
\FOR{$t = 1$ \textbf{to} $T$}
    \STATE Sample $\tilde{\theta} \sim \mathcal{N}(A^{-1}b,\, A^{-1})$ \hfill\textcolor{gray}{\textit{// diversity-aware bandit item selection}}
    \STATE $c \leftarrow \arg\max_{c \in \mathcal{C}} \big[\psi(c)^\top \tilde{\theta} + \beta\, d(c, \mathcal{R})\big]$
    \STATE $h \leftarrow \text{LLM}_{\text{mut}}(\pi, c, \text{SampleParents}(\mathcal{A}, m))$ \hfill\textcolor{gray}{\textit{// few-shot hypothesis generation}}
    \STATE $x_{\text{new}} \leftarrow (c, h, \mathrm{Embed}(c,h))$ \hfill\textcolor{gray}{\textit{// local novelty + ranking archive update}}
    \STATE Obtain $q(x_{\text{new}})$; set $r\leftarrow r(x_{\text{new}})$ \hfill\textcolor{gray}{\textit{// score scaled to $[0,1]$}}
    \STATE $\mathcal{N}_k \leftarrow k\text{NN}(x_{\text{new}}, \mathcal{A})$
    \STATE $\texttt{novel} \leftarrow \text{LLM}_{\text{novelty}}(\pi, x_{\text{new}}, \mathcal{N}_k)$ \hfill\textcolor{gray}{\textit{// novelty first}}
    \IF{$\texttt{novel}$}
        \STATE Add $x_{\text{new}}$ to $\mathcal{A}$ \hfill\textcolor{gray}{\textit{// admitted; no comparison call}}
    \ELSIF{$\mathrm{Better}_{\pi}(x_{\text{new}},\ \text{nearest neighbor})$}
        \STATE Replace nearest neighbor with $x_{\text{new}}$ \hfill\textcolor{gray}{\textit{// blind pairwise duel}}
    \ENDIF
    \STATE $A \leftarrow A + \sigma^{-2}\psi(c)\psi(c)^\top$,\ \ $b \leftarrow b + \sigma^{-2}\,r\,\psi(c)$ \hfill\textcolor{gray}{\textit{// update even if discarded}}
\ENDFOR
\STATE \textbf{return} $\mathcal{A}$
\end{algorithmic}
\end{algorithm*}

\subsection{Implementation details}
\label{sec:implementation}
To speed up experiments, we run the loop in parallel batches where workers propose and evaluate candidates against a snapshot of the current archive. To select a batch of $B$ items, we extend the single-draw rule of Section~\ref{sec:full-method}: we take $B$ independent posterior draws $\tilde{\theta}_1,\dots,\tilde{\theta}_B$, and for each $\tilde{\theta}_i$ select the candidate maximizing $s(c)$ while excluding items already chosen in the batch and adding them to $\mathcal{R}$, so the diversity bonus discourages near-duplicates within a batch as well as against the archive. After each batch, we reconcile results by filtering near-duplicate additions (above a cosine-similarity threshold) and resolving replacement collisions (multiple workers targeting the same neighbor). The bandit uses prior precision $\lambda=1$ and noise standard deviation $\sigma=0.1$, the novelty check uses $k=3$ nearest neighbors, and each run starts from $10$ seed entries.

\section{Experiments}
\label{sec:experiments}

We evaluate the three variants (Novelty, Novelty+Evo, QD-Rec; Table~\ref{tab:method-variants}) across four domains.

\subsection{Setup}
\label{sec:setup}
\paragraph{Common protocol.}
Our main experiment uses \texttt{claude-sonnet-5} as the hypothesis proposer; we additionally report a weaker-proposer ablation that swaps in \texttt{claude-haiku-4-5}. The judge is \texttt{gpt-5.6-terra} in \emph{every} configuration and domain, a different model family from both proposers, to prevent self-preference bias \citep{wataoka2025neuripsws}. Proposer and judge both run with reasoning effort set to \texttt{high}. The weak-proposer ablation model does not support the effort parameter and runs without it or extended thinking. We leave the temperature setting to the provider default for \texttt{claude-haiku-4-5}. Each configuration runs $1{,}000$ evaluations (proposal--judge cycles) per method in all four domains, executed in generations of up to $20$ parallel evaluations, with the archive reconciled between generations (Section~\ref{sec:implementation}). Novelty+Evo and QD-Rec condition each proposal on $m=3$ archive exemplars. Runs are repeated for five random seeds.
Each run begins from a seed archive under the same profile $\pi$; the seed controls the pipeline's own sampling randomness (parent selection, item ordering) while LLM sampling stays at the provider default, so across-seed spread measures the stochasticity of the full pipeline.

We embed archive entries and candidate items with OpenAI's \texttt{text-embedding-3-small} at $512$ output dimensions, L2-normalized, and use cosine similarity for $k$-nearest-neighbor retrieval.

\paragraph{Methods.}
We evaluate the proposed method QD-Rec and its two ablations (Novelty, Novelty+Evo) under matched evaluation budgets and shared hyperparameters within each domain. Table~\ref{tab:method-variants} summarizes the three variants as an ``ingredients'' view: all share the archive mechanics; Novelty+Evo adds the evolutionary few-shot operator, and QD-Rec additionally adds diversity-aware bandit item selection.

\begin{table}[!t]
\centering
\small
\begin{tabular}{lccc}
\toprule
 & Local novelty filter & Evolutionary operator & Bandit item selection \\
\midrule
Novelty & \textcolor{green!50!black}{$\checkmark$} & \textcolor{red!75!black}{$\times$} & \textcolor{red!75!black}{$\times$} \\
Novelty+Evo & \textcolor{green!50!black}{$\checkmark$} & \textcolor{green!50!black}{$\checkmark$} & \textcolor{red!75!black}{$\times$} \\
QD-Rec (full) & \textcolor{green!50!black}{$\checkmark$} & \textcolor{green!50!black}{$\checkmark$} & \textcolor{green!50!black}{$\checkmark$} \\
\bottomrule
\end{tabular}
\caption{QD-Rec and its two ablations. All three use the same archive update (Section~\ref{sec:archive-update}).}
\label{tab:method-variants}
\end{table}

\paragraph{Domains.}
\label{sec:domains}
We instantiate the framework on four fixed-pool domains, each realizing the hypothesis $h$ in its native genre. The first, Goodreads, is a \emph{grounding check} whose quality signal is a real human rating rather than an LLM judge; the other three are heterogeneous applied settings. Full dataset construction (sources, filtering, pool sizes, and the specific profiles used) is in Appendix~\ref{app:datasets}.

\setlength{\leftmargini}{1.4em}
\begin{itemize}
\setlength{\itemsep}{2pt}\setlength{\parskip}{0pt}\setlength{\topsep}{2pt}
\item \textbf{Goodreads.} The pool $\mathcal{C}$ is the children's books a reader has rated \citep{wan2018goodreads}; the profile $\pi$ is that reader (summarized from past reviews); $h$ is a reading memo on whether they would enjoy a book. Uniquely, quality is \emph{not} an LLM judge but the reader's own $1$--$5$ rating, held out of all prompts and used only as the reward and replacement signal (Appendix~\ref{app:datasets}), testing whether QD still yields meaningful quality--diversity tradeoffs under a real human signal. This Quality measure evaluates the book for the reader, not the generated memo: changing the memo cannot change the rating, although it can affect novelty and archive geometry.
\item \textbf{M\&A target discovery.} $\mathcal{C}$ is a universe of 6{,}332 US companies built from LSEG Workspace data (business descriptions, industry taxonomy, financials). The business descriptions average approximately 123 words. The profile $\pi$ is an acquirer (Walt Disney or JPMorgan Chase), and $h$ is an acquisition thesis (e.g.\ capability acquisition vs.\ market access vs.\ vertical integration). Deal sourcing motivates it: teams want a \emph{diverse set of strategic theses} rather than a single shortlist.
\item \textbf{Marketing.} $\mathcal{C}$ is 10{,}000 synthetic customer profiles from California, sampled from NVIDIA Nemotron-Personas-USA \citep{nemotron-personas-usa}. The customer profiles average approximately 685 words. The profile $\pi$ is a product (a desktop 3D printer or an online upskilling subscription), and $h$ is a targeting hypothesis explaining why a profile is an attractive target for $\pi$, yielding a map of distinct targeting angles rather than a ranked list.
\item \textbf{Paper-abstract exploration.} $\mathcal{C}$ is the titles and abstracts of 17{,}511 papers accepted at ICLR 2026, ICML 2026 and NeurIPS 2025 \citep{paperlists}, $\pi$ a researcher profile (an open-endedness researcher or an applied researcher at an AI startup), and $h$ a research memo linking a paper to $\pi$. The scale of the AI literature motivates it: researchers need \emph{diverse}, well-justified reading directions under limited time. Candidate embeddings for bandit item selection are precomputed from title+abstract; archive embeddings are computed from title+abstract+memo.
\end{itemize}

\needspace{4\baselineskip}
\paragraph{Evaluation.}
Four metrics summarize a run's final active archive.

\begin{itemize}
\setlength{\itemsep}{2pt}\setlength{\parskip}{0pt}\setlength{\topsep}{2pt}
\item Quality is the mean pointwise score of its entries: the judge score (0--10), or in Goodreads the reader's real star rating (1--5).
\item Diversity is one minus the mean pairwise cosine similarity between archive-entry embeddings, computed in the original embedding space (the complement of the mean pairwise similarity used to quantify homogenization by \citealp{jiang2025artificial}).
\item Coverage is the fraction of occupied cells on a $25{\times}25$ grid over a 2D PCA projection (the maps of Figure~\ref{fig:main-map}). For each profile, proposer, and trial seed, we fit PCA by full SVD on the union of all three variants' active archives and use shared axis bounds padded by 1\% on each side. Coverage is comparable across methods within that grid; separate profiles, proposers, and seeds use separate grids.
\item Archive size $n$ is the number of entries that survived admission.
\end{itemize}

\paragraph{Online evaluation protocol.}
We score each recommendation during the search. QD-Rec uses these scores to update its bandit and select subsequent candidates. Final Quality is the mean of the scores already assigned to the recommendations retained in the archive.

\subsection{Results}
\label{sec:results}

Table~\ref{tab:exp2-main} reports our main results: mean Quality and semantic Diversity of the three variants (Novelty, Novelty+Evo, QD-Rec) on both profiles of every domain, under both proposers.

\begin{table*}[!tp]
  \centering
  \setlength{\tabcolsep}{4pt}
  \resizebox{\linewidth}{!}{%
  \begin{tabular}{llll cccc cccc cccc}
  \toprule
  Domain & Profile & Proposer & Judge &
  \multicolumn{4}{c}{Novelty} &
  \multicolumn{4}{c}{Novelty+Evo} &
  \multicolumn{4}{c}{QD-Rec} \\
  \cmidrule(lr){5-8}\cmidrule(lr){9-12}\cmidrule(lr){13-16}
   &  &  &  & Qual. & Div. & Cov. & $n$ & Qual. & Div. & Cov. & $n$ & Qual. & Div. & Cov. & $n$ \\
  \midrule
  Goodreads$^{\dagger}$ & series/YA & Sonnet 5 & human & \cellcolor{abccol}\textbf{4.17{\scriptsize$\pm$0.01}} & .463{\scriptsize$\pm$.001} & .26{\scriptsize$\pm$.00} & 222{\scriptsize$\pm$5} & \cellcolor{abccol}\textbf{4.15{\scriptsize$\pm$0.01}} & \cellcolor{abccol}\textbf{.480{\scriptsize$\pm$.001}} & \cellcolor{abccol}\textbf{.27{\scriptsize$\pm$.00}} & \cellcolor{abccol}\textbf{226{\scriptsize$\pm$5}} & \cellcolor{abccol}\textbf{4.15{\scriptsize$\pm$0.02}} & \cellcolor{abccol}\textbf{.480{\scriptsize$\pm$.001}} & \cellcolor{abccol}\textbf{.28{\scriptsize$\pm$.01}} & \cellcolor{abccol}\textbf{244{\scriptsize$\pm$3}} \\
   &  & Haiku 4.5 & human & \cellcolor{abccol}\textbf{4.09{\scriptsize$\pm$0.01}} & .480{\scriptsize$\pm$.001} & .26{\scriptsize$\pm$.00} & 233{\scriptsize$\pm$2} & 4.08{\scriptsize$\pm$0.01} & \cellcolor{abccol}\textbf{.493{\scriptsize$\pm$.001}} & \cellcolor{abccol}\textbf{.29{\scriptsize$\pm$.00}} & 254{\scriptsize$\pm$3} & \cellcolor{abccol}\textbf{4.12{\scriptsize$\pm$0.01}} & \cellcolor{abccol}\textbf{.492{\scriptsize$\pm$.002}} & \cellcolor{abccol}\textbf{.28{\scriptsize$\pm$.01}} & \cellcolor{abccol}\textbf{257{\scriptsize$\pm$3}} \\
   & picture-book & Sonnet 5 & human & \cellcolor{abccol}\textbf{4.19{\scriptsize$\pm$0.01}} & .475{\scriptsize$\pm$.002} & .32{\scriptsize$\pm$.00} & 297{\scriptsize$\pm$6} & \cellcolor{abccol}\textbf{4.17{\scriptsize$\pm$0.02}} & \cellcolor{abccol}\textbf{.481{\scriptsize$\pm$.002}} & .33{\scriptsize$\pm$.01} & 318{\scriptsize$\pm$6} & 4.12{\scriptsize$\pm$0.02} & \cellcolor{abccol}\textbf{.485{\scriptsize$\pm$.001}} & \cellcolor{abccol}\textbf{.34{\scriptsize$\pm$.01}} & \cellcolor{abccol}\textbf{336{\scriptsize$\pm$7}} \\
   &  & Haiku 4.5 & human & \cellcolor{abccol}\textbf{4.14{\scriptsize$\pm$0.01}} & .489{\scriptsize$\pm$.001} & .32{\scriptsize$\pm$.01} & 320{\scriptsize$\pm$2} & \cellcolor{abccol}\textbf{4.15{\scriptsize$\pm$0.02}} & .497{\scriptsize$\pm$.003} & .33{\scriptsize$\pm$.01} & 337{\scriptsize$\pm$4} & \cellcolor{abccol}\textbf{4.11{\scriptsize$\pm$0.02}} & \cellcolor{abccol}\textbf{.506{\scriptsize$\pm$.001}} & \cellcolor{abccol}\textbf{.35{\scriptsize$\pm$.01}} & \cellcolor{abccol}\textbf{367{\scriptsize$\pm$4}} \\
  \addlinespace
  M\&A & Walt Disney Co & Sonnet 5 & GPT-5.6 & 2.40{\scriptsize$\pm$0.12} & .330{\scriptsize$\pm$.001} & .12{\scriptsize$\pm$.00} & 85{\scriptsize$\pm$1} & 2.51{\scriptsize$\pm$0.07} & \cellcolor{abccol}\textbf{.353{\scriptsize$\pm$.001}} & .18{\scriptsize$\pm$.01} & 135{\scriptsize$\pm$5} & \cellcolor{abccol}\textbf{2.91{\scriptsize$\pm$0.05}} & \cellcolor{abccol}\textbf{.352{\scriptsize$\pm$.001}} & \cellcolor{abccol}\textbf{.26{\scriptsize$\pm$.00}} & \cellcolor{abccol}\textbf{214{\scriptsize$\pm$2}} \\
   &  & Haiku 4.5 & GPT-5.6 & 2.12{\scriptsize$\pm$0.05} & \cellcolor{abccol}\textbf{.326{\scriptsize$\pm$.002}} & .19{\scriptsize$\pm$.01} & 143{\scriptsize$\pm$4} & 2.11{\scriptsize$\pm$0.04} & \cellcolor{abccol}\textbf{.322{\scriptsize$\pm$.002}} & .23{\scriptsize$\pm$.01} & 188{\scriptsize$\pm$4} & \cellcolor{abccol}\textbf{2.55{\scriptsize$\pm$0.03}} & .318{\scriptsize$\pm$.001} & \cellcolor{abccol}\textbf{.27{\scriptsize$\pm$.00}} & \cellcolor{abccol}\textbf{262{\scriptsize$\pm$5}} \\
   & JPMorgan & Sonnet 5 & GPT-5.6 & 2.88{\scriptsize$\pm$0.04} & .332{\scriptsize$\pm$.001} & .13{\scriptsize$\pm$.01} & 95{\scriptsize$\pm$3} & 2.54{\scriptsize$\pm$0.08} & \cellcolor{abccol}\textbf{.363{\scriptsize$\pm$.003}} & .15{\scriptsize$\pm$.01} & 108{\scriptsize$\pm$4} & \cellcolor{abccol}\textbf{3.04{\scriptsize$\pm$0.04}} & .353{\scriptsize$\pm$.001} & \cellcolor{abccol}\textbf{.23{\scriptsize$\pm$.00}} & \cellcolor{abccol}\textbf{183{\scriptsize$\pm$5}} \\
   &  & Haiku 4.5 & GPT-5.6 & 2.25{\scriptsize$\pm$0.04} & \cellcolor{abccol}\textbf{.331{\scriptsize$\pm$.001}} & .20{\scriptsize$\pm$.00} & 150{\scriptsize$\pm$3} & 2.17{\scriptsize$\pm$0.04} & .324{\scriptsize$\pm$.003} & .22{\scriptsize$\pm$.01} & 166{\scriptsize$\pm$9} & \cellcolor{abccol}\textbf{2.71{\scriptsize$\pm$0.02}} & .316{\scriptsize$\pm$.001} & \cellcolor{abccol}\textbf{.28{\scriptsize$\pm$.00}} & \cellcolor{abccol}\textbf{238{\scriptsize$\pm$7}} \\
  \addlinespace
  ML Papers & OpenEndedness & Sonnet 5 & GPT-5.6 & 4.96{\scriptsize$\pm$0.02} & .469{\scriptsize$\pm$.001} & .43{\scriptsize$\pm$.01} & 578{\scriptsize$\pm$5} & 5.26{\scriptsize$\pm$0.02} & .479{\scriptsize$\pm$.002} & \cellcolor{abccol}\textbf{.47{\scriptsize$\pm$.01}} & 690{\scriptsize$\pm$8} & \cellcolor{abccol}\textbf{5.32{\scriptsize$\pm$0.02}} & \cellcolor{abccol}\textbf{.501{\scriptsize$\pm$.001}} & .44{\scriptsize$\pm$.01} & \cellcolor{abccol}\textbf{730{\scriptsize$\pm$8}} \\
   &  & Haiku 4.5 & GPT-5.6 & 4.26{\scriptsize$\pm$0.02} & .489{\scriptsize$\pm$.001} & \cellcolor{abccol}\textbf{.39{\scriptsize$\pm$.01}} & \cellcolor{abccol}\textbf{468{\scriptsize$\pm$9}} & \cellcolor{abccol}\textbf{5.03{\scriptsize$\pm$0.07}} & .504{\scriptsize$\pm$.002} & .26{\scriptsize$\pm$.02} & 247{\scriptsize$\pm$31} & \cellcolor{abccol}\textbf{4.87{\scriptsize$\pm$0.02}} & \cellcolor{abccol}\textbf{.523{\scriptsize$\pm$.002}} & .30{\scriptsize$\pm$.01} & 353{\scriptsize$\pm$34} \\
   & AI Startup & Sonnet 5 & GPT-5.6 & \cellcolor{abccol}\textbf{6.21{\scriptsize$\pm$0.01}} & .495{\scriptsize$\pm$.001} & .46{\scriptsize$\pm$.00} & 658{\scriptsize$\pm$7} & \cellcolor{abccol}\textbf{6.24{\scriptsize$\pm$0.03}} & .502{\scriptsize$\pm$.001} & \cellcolor{abccol}\textbf{.48{\scriptsize$\pm$.00}} & 737{\scriptsize$\pm$8} & \cellcolor{abccol}\textbf{6.24{\scriptsize$\pm$0.02}} & \cellcolor{abccol}\textbf{.521{\scriptsize$\pm$.001}} & .45{\scriptsize$\pm$.01} & \cellcolor{abccol}\textbf{789{\scriptsize$\pm$8}} \\
   &  & Haiku 4.5 & GPT-5.6 & \cellcolor{abccol}\textbf{5.81{\scriptsize$\pm$0.02}} & .516{\scriptsize$\pm$.001} & \cellcolor{abccol}\textbf{.46{\scriptsize$\pm$.01}} & 616{\scriptsize$\pm$4} & \cellcolor{abccol}\textbf{5.87{\scriptsize$\pm$0.03}} & .507{\scriptsize$\pm$.001} & \cellcolor{abccol}\textbf{.47{\scriptsize$\pm$.01}} & 674{\scriptsize$\pm$12} & \cellcolor{abccol}\textbf{5.83{\scriptsize$\pm$0.02}} & \cellcolor{abccol}\textbf{.526{\scriptsize$\pm$.002}} & .43{\scriptsize$\pm$.01} & \cellcolor{abccol}\textbf{720{\scriptsize$\pm$6}} \\
  \addlinespace
  Marketing & 3D printer & Sonnet 5 & GPT-5.6 & 6.81{\scriptsize$\pm$0.04} & .219{\scriptsize$\pm$.001} & .13{\scriptsize$\pm$.01} & 96{\scriptsize$\pm$4} & 6.73{\scriptsize$\pm$0.01} & .237{\scriptsize$\pm$.001} & \cellcolor{abccol}\textbf{.18{\scriptsize$\pm$.01}} & \cellcolor{abccol}\textbf{156{\scriptsize$\pm$5}} & \cellcolor{abccol}\textbf{7.05{\scriptsize$\pm$0.05}} & \cellcolor{abccol}\textbf{.243{\scriptsize$\pm$.001}} & \cellcolor{abccol}\textbf{.18{\scriptsize$\pm$.00}} & \cellcolor{abccol}\textbf{164{\scriptsize$\pm$2}} \\
   &  & Haiku 4.5 & GPT-5.6 & \cellcolor{abccol}\textbf{6.65{\scriptsize$\pm$0.05}} & .190{\scriptsize$\pm$.003} & .11{\scriptsize$\pm$.00} & 80{\scriptsize$\pm$3} & 6.29{\scriptsize$\pm$0.02} & \cellcolor{abccol}\textbf{.202{\scriptsize$\pm$.002}} & \cellcolor{abccol}\textbf{.14{\scriptsize$\pm$.00}} & \cellcolor{abccol}\textbf{110{\scriptsize$\pm$3}} & \cellcolor{abccol}\textbf{6.83{\scriptsize$\pm$0.03}} & \cellcolor{abccol}\textbf{.199{\scriptsize$\pm$.001}} & \cellcolor{abccol}\textbf{.13{\scriptsize$\pm$.00}} & \cellcolor{abccol}\textbf{104{\scriptsize$\pm$3}} \\
   & Upskilling & Sonnet 5 & GPT-5.6 & 6.47{\scriptsize$\pm$0.08} & .225{\scriptsize$\pm$.001} & .09{\scriptsize$\pm$.00} & 61{\scriptsize$\pm$1} & \cellcolor{abccol}\textbf{6.70{\scriptsize$\pm$0.06}} & \cellcolor{abccol}\textbf{.230{\scriptsize$\pm$.002}} & \cellcolor{abccol}\textbf{.10{\scriptsize$\pm$.00}} & \cellcolor{abccol}\textbf{72{\scriptsize$\pm$3}} & \cellcolor{abccol}\textbf{6.89{\scriptsize$\pm$0.06}} & \cellcolor{abccol}\textbf{.235{\scriptsize$\pm$.001}} & \cellcolor{abccol}\textbf{.11{\scriptsize$\pm$.00}} & \cellcolor{abccol}\textbf{82{\scriptsize$\pm$3}} \\
   &  & Haiku 4.5 & GPT-5.6 & \cellcolor{abccol}\textbf{6.47{\scriptsize$\pm$0.03}} & .221{\scriptsize$\pm$.001} & \cellcolor{abccol}\textbf{.08{\scriptsize$\pm$.01}} & \cellcolor{abccol}\textbf{69{\scriptsize$\pm$4}} & \cellcolor{abccol}\textbf{6.35{\scriptsize$\pm$0.12}} & \cellcolor{abccol}\textbf{.238{\scriptsize$\pm$.005}} & \cellcolor{abccol}\textbf{.08{\scriptsize$\pm$.00}} & 58{\scriptsize$\pm$3} & \cellcolor{abccol}\textbf{6.62{\scriptsize$\pm$0.10}} & \cellcolor{abccol}\textbf{.245{\scriptsize$\pm$.002}} & \cellcolor{abccol}\textbf{.07{\scriptsize$\pm$.00}} & \cellcolor{abccol}\textbf{54{\scriptsize$\pm$3}} \\
  \bottomrule
  \end{tabular}}
  \caption{Main results: mean $\pm$ standard error over five seeds (metrics in Section~\ref{sec:setup}; $\dagger$: Goodreads Quality is a human star rating from $1$ to $5$). Per metric, bold text with blue shading marks the best mean and every cell not significantly different from it (paired two-sided $t$-test across seeds, $p \ge 0.05$, uncorrected; the three variants share each seed's initial archive). A single highlighted cell is therefore significantly better than both other variants.}
  \label{tab:exp2-main}
\end{table*}

\paragraph{Main results.}
QD-Rec gives the best mean Quality in most LLM-judged settings (Table~\ref{tab:exp2-main}), often significantly better than both ablations, and it is never significantly worse than the best variant. The gains are largest on M\&A. QD-Rec also builds the largest archive in most settings and often has the best MAP-Elites coverage, so its higher Quality does not come from a smaller, more selective archive. Its semantic Diversity is the best or tied for best in every domain except M\&A, where it is slightly lower. On Goodreads, where Quality is the reader's own star rating, the three variants reach similar mean ratings. There, QD-Rec keeps this rating while matching or beating both ablations on archive size, Diversity and coverage.

\paragraph{Cost.}
We measure cost as the API spend of a run in US dollars at list prices. It covers every model call: the proposer, the quality judge, the novelty check, the duel, and the embeddings. The three variants make the same calls for each of the $1{,}000$ evaluations. They differ only in the proposer prompt, which includes three archive exemplars in Novelty+Evo and QD-Rec, and in how often a duel is needed. The bandit runs locally and adds no API cost. The runs did not log token usage, so we estimate the cost of each call type by re-sending a sample of $840$ archived calls with the same models and settings. We then multiply by the exact number of calls in each run. QD-Rec costs the same as Novelty+Evo, and both cost about $1.3\times$ as much as Novelty. More than $90\%$ of the difference comes from the proposer, because the exemplars roughly triple its input. A single run costs between about \$13 (Novelty on Goodreads with the \texttt{claude-haiku-4-5} proposer) and \$62 (QD-Rec on Marketing with the \texttt{claude-sonnet-5} proposer). Goodreads is the cheapest domain because its ratings replace the LLM judge and the duel.

  \begin{figure*}[!t]
      \centering
      {\small M\&A: Walt Disney Co}\\[1pt]
      \includegraphics[width=0.44\textwidth]{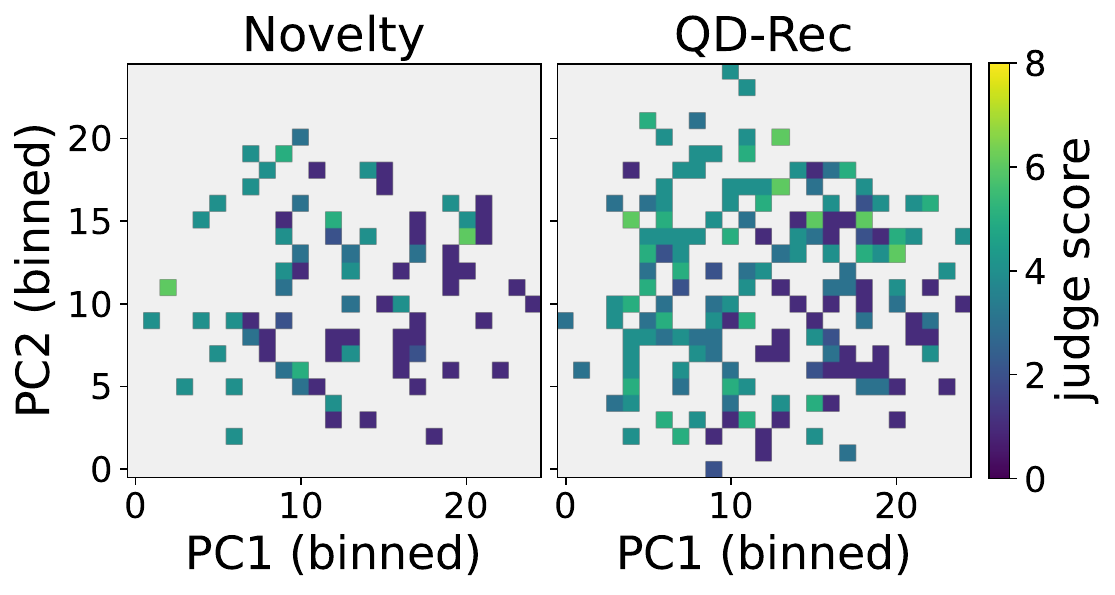}\hfill
      \includegraphics[width=0.25\textwidth]{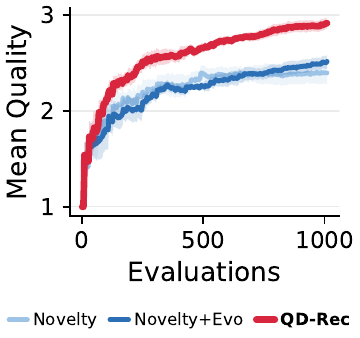}\hfill
      \includegraphics[width=0.25\textwidth]{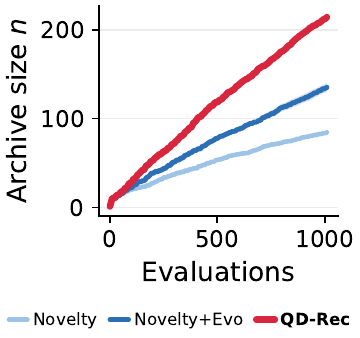}\\[4pt]
      {\small ML Papers: OpenEndedness}\\[1pt]
      \includegraphics[width=0.44\textwidth]{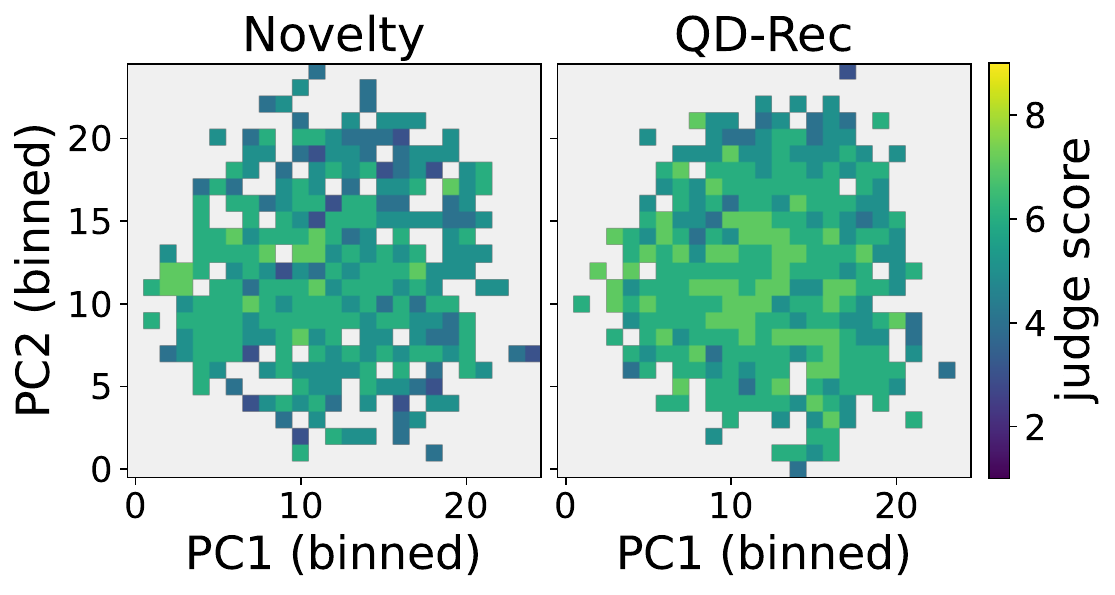}\hfill
      \includegraphics[width=0.25\textwidth]{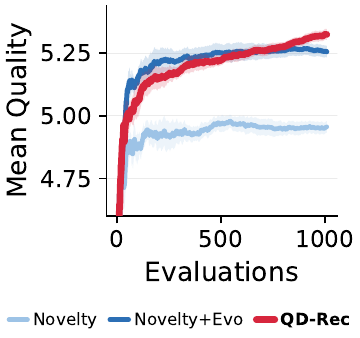}\hfill
      \includegraphics[width=0.25\textwidth]{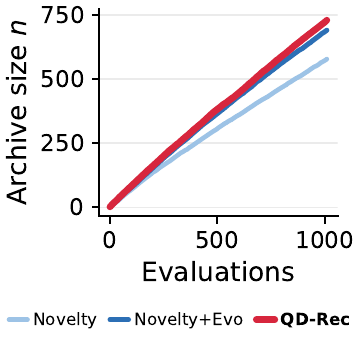}\\[4pt]
      {\small Marketing: 3D printer}\\[1pt]
      \subfigure[Archive map]{\includegraphics[width=0.44\textwidth]{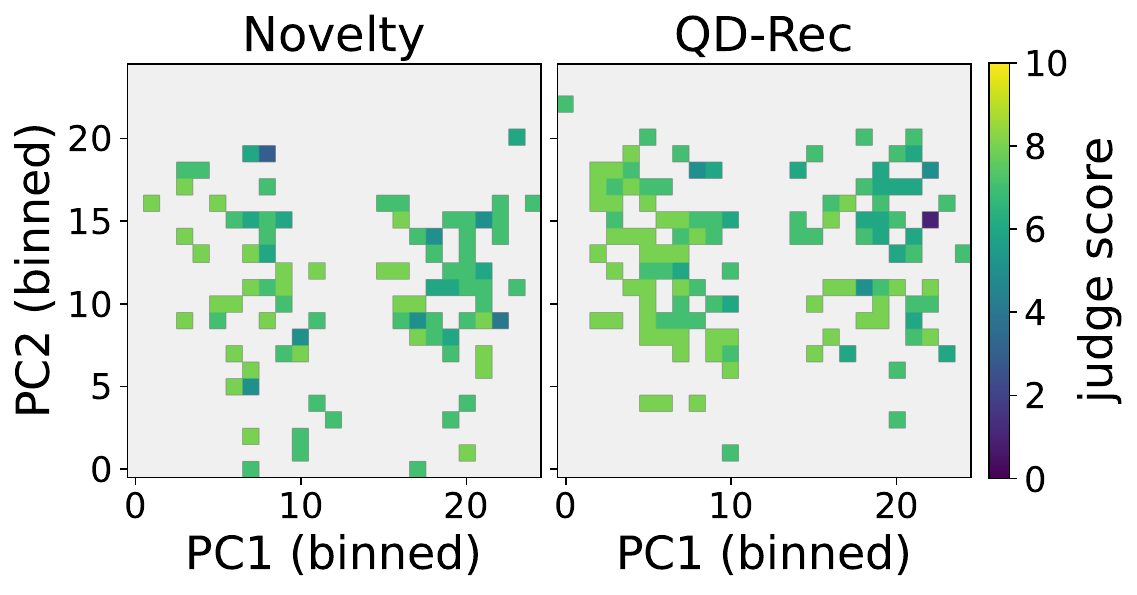}\label{fig:main-map}}
      \hfill
      \subfigure[Quality over the run]{\includegraphics[width=0.25\textwidth]{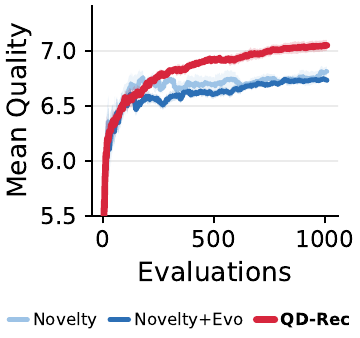}\label{fig:main-quality}}
      \hfill
      \subfigure[Archive size over the run]{\includegraphics[width=0.25\textwidth]{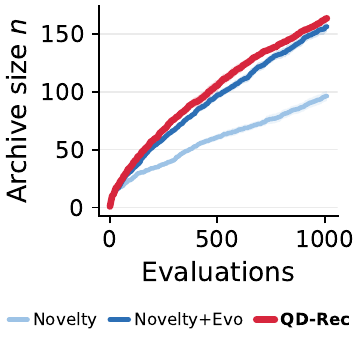}\label{fig:main-size}}

      \caption{Three profiles, one per row (M\&A: Walt Disney Co; ML Papers: OpenEndedness; Marketing: 3D printer). \textbf{(a)}~MAP-Elites maps of Novelty and QD-Rec (best quality per cell of the Coverage grid, Section~\ref{sec:setup}; seed~$2$). \textbf{(b,c)}~Mean Quality and size $n$ of the current active archive after every evaluation (\texttt{claude-sonnet-5} proposer, mean $\pm$ standard error over five seeds). All eight scenarios are in the appendix: maps in Figure~\ref{fig:maps-extra} and curves over the run in Figure~\ref{fig:qd-trajectory-extra}.}
      \label{fig:main}
      \end{figure*}

\paragraph{Qualitative maps.}
The left column of Figure~\ref{fig:main} shows MAP-Elites-style heatmaps for Novelty and QD-Rec on one profile per domain. In each, QD-Rec covers more of the space than Novelty, most visibly for Disney and the 3D printer. Maps for all eight scenarios, including Goodreads and Novelty+Evo, are in Appendix~\ref{app:results-table}, and example recommendations are in Appendix~\ref{app:examples}.

\paragraph{Comparison to a no-learning baseline.}
To ask whether the online learning and archive machinery earns its cost, we compare against two training-free re-rankers, relevance-MMR \citep{carbonell1998use} and a quality-weighted DPP with greedy MAP inference \citep{chen2018dpp}, on Goodreads, where Quality is a real human star rating. Sweeping each baseline's relevance--diversity weight $\gamma$,no setting of either reaches any QD variant on Quality, and they approach QD-Rec's semantic Diversity only at their most diversity-favoring settings. The baselines see no ratings while the QD variants query up to $1{,}000$, so the gap reflects rating feedback and the archive together, not item selection alone. Full setup and the per-setting table are in Appendix~\ref{app:mmr}.

\paragraph{Quality and archive size over the run.}
Figures~\ref{fig:main-quality} and~\ref{fig:main-size} plot the current archive's mean Quality and size $n$ after every evaluation, replayed from each run's add/replace/discard history (all eight scenarios in Figure~\ref{fig:qd-trajectory-extra}). QD-Rec's Quality keeps rising through the run and ends highest, so its advantage builds up during the run. Archive size $n$ grows throughout the run as novel hypotheses are added.

\begin{table}[t]
  \centering
  \setlength{\tabcolsep}{4pt}
  \resizebox{0.8\linewidth}{!}{%
  \begin{tabular}{ll cccc cccc cccc}
  \toprule
  Domain & Profile &
  \multicolumn{4}{c}{Novelty} &
  \multicolumn{4}{c}{Novelty+Evo} &
  \multicolumn{4}{c}{QD-Rec} \\
  \cmidrule(lr){3-6}\cmidrule(lr){7-10}\cmidrule(lr){11-14}
   &  & Qual. & Div. & Cov. & $n$ & Qual. & Div. & Cov. & $n$ & Qual. & Div. & Cov. & $n$ \\
  \midrule
  Goodreads$^{\dagger}$ & series/YA & 4.12 & .496 & .27 & 282 & 4.10 & .523 & \cellcolor{abccol}\textbf{.28} & 280 & \cellcolor{abccol}\textbf{4.14} & \cellcolor{abccol}\textbf{.533} & \cellcolor{abccol}\textbf{.28} & \cellcolor{abccol}\textbf{287} \\
  M\&A & Walt Disney Co & 2.05 & \cellcolor{abccol}\textbf{.520} & .18 & 141 & 1.86 & .512 & .26 & 209 & \cellcolor{abccol}\textbf{2.30} & .518 & \cellcolor{abccol}\textbf{.29} & \cellcolor{abccol}\textbf{270} \\
  ML Papers & OpenEndedness & 4.25 & .541 & \cellcolor{abccol}\textbf{.44} & \cellcolor{abccol}\textbf{538} & \cellcolor{abccol}\textbf{5.05} & .567 & .39 & 402 & 4.86 & \cellcolor{abccol}\textbf{.623} & .34 & 425 \\
  Marketing & 3D printer & 6.72 & .444 & .14 & 99 & 6.44 & \cellcolor{abccol}\textbf{.474} & \cellcolor{abccol}\textbf{.19} & \cellcolor{abccol}\textbf{144} & \cellcolor{abccol}\textbf{6.89} & .470 & .17 & 129 \\
  \bottomrule
  \end{tabular}}
  \caption{Embedding-model sensitivity: one profile per domain re-run with \texttt{Qwen3-Embedding-0.6B} embeddings (\texttt{claude-haiku-4-5} proposer, seed $0$, one run). Bold with blue shading marks the best cell per metric; with a single run, no significance test is applied.}
  \label{tab:emb-sens}
\end{table}

\paragraph{Embedding-model sensitivity.}
The embedding model defines the novelty neighborhoods, the Diversity metric, the MAP-Elites grid and the bandit's item features. We therefore re-run one profile per domain with a different embedding model, \texttt{Qwen3-Embedding-0.6B} truncated to the same $512$ dimensions (\texttt{claude-haiku-4-5} proposer, seed $0$, one run with batches of $16$ instead of $20$, due to the compute limits of the local machine that runs the embedding model; Table~\ref{tab:emb-sens}). The best variant on Quality and on archive size is the same as with the original embeddings on all four profiles. Diversity is higher for every variant, because the Qwen embeddings are spread more evenly. Diversity and Coverage can therefore be compared only within one embedding space, while the Quality and archive-size orderings are unchanged.

\section{Conclusion}
We introduced QD-Rec, an archive-based quality--diversity method for open-ended recommendation, where the items are chosen with contextual bandits, and the reasons for recommending them are searched open-endedly. Across four domains, QD-Rec gave the best mean Quality in most settings and usually built the largest archive. As a first work along this line, we kept it relatively simple, and several directions follow.

First, our bandit uses Bayesian linear regression to model reward from fixed item embeddings. Neural contextual bandits \citep{zhou2020icml,zhang2021iclr} could capture nonlinear relationships between these embeddings and reward. More radically, the bandit could be discarded altogether in favor of a coding agent that reads the archive directly, reasons over which regions look exhausted and which look promising, and writes its own selection rule, replacing a fixed parametric reward model with a policy that can be inspected and revised in natural language. Second, exemplars are sampled uniformly from the archive, and could instead be selected by quality, recency, or how often they have already served as exemplars \citep{earle2026gecco,zhang2026iclr}. Bandits could also be used for parent selection, in addition to item selection \citep{sfikas2021gecco,gallotta2024tg}. Third, this work focused on the \emph{outer loop}: which item to evaluate next, and which pairs to retain. The inner hypothesis generator was a single prompted LLM call, and could instead be a full research agent \citep{shang2025neurips,lin2026arxiv}; we kept it lightweight to show that the outer loop is itself a meaningful lever, and expect the two to compose.

\bibliographystyle{plainnat}
\subsubsection*{Acknowledgments}
We thank David Ha and an anonymous friend for insightful suggestions and discussions, Artsem Zhyvalkouski for help with data preparation, and Daichi Tazawa for help with coordination.

\bibliography{references}

\begin{thebibliography}{53}
\providecommand{\natexlab}[1]{#1}
\providecommand{\url}[1]{\texttt{#1}}
\expandafter\ifx\csname urlstyle\endcsname\relax
  \providecommand{\doi}[1]{doi: #1}\else
  \providecommand{\doi}{doi: \begingroup \urlstyle{rm}\Url}\fi

\bibitem[Agarwal et~al.(2009)Agarwal, Chen, and Elango]{agarwal2009icdm}
Deepak Agarwal, Bee-Chung Chen, and Pradheep Elango.
\newblock Explore/exploit schemes for web content optimization.
\newblock In \emph{ICDM}, 2009.

\bibitem[Aggarwal(2016)]{aggarwal2016recommender}
Charu~C. Aggarwal.
\newblock \emph{Recommender Systems: The Textbook}.
\newblock Springer, 2016.

\bibitem[Agrawal and Goyal(2013)]{agrawal2013icml}
Shipra Agrawal and Navin Goyal.
\newblock Thompson sampling for contextual bandits with linear payoffs.
\newblock In \emph{ICML}, 2013.

\bibitem[Bahlous-Boldi et~al.(2025)Bahlous-Boldi, Faldor, Grillotti,
  Janmohamed, Coiffard, Spector, and Cully]{bahlous2025dominated}
Ryan Bahlous-Boldi, Maxence Faldor, Luca Grillotti, Hannah Janmohamed, Lisa
  Coiffard, Lee Spector, and Antoine Cully.
\newblock Dominated novelty search: Rethinking local competition in
  quality-diversity.
\newblock In \emph{GECCO}, 2025.

\bibitem[Bradley et~al.(2024)Bradley, Dai, Teufel, Zhang, Oostermeijer,
  Bellagente, Clune, Stanley, Schott, and Lehman]{bradley2023quality}
Herbie Bradley, Andrew Dai, Hannah Teufel, Jenny Zhang, Koen Oostermeijer,
  Marco Bellagente, Jeff Clune, Kenneth Stanley, Gr{\'e}gory Schott, and Joel
  Lehman.
\newblock Quality-diversity through {AI} feedback.
\newblock In \emph{ICLR}, 2024.

\bibitem[Carbonell and Goldstein(1998)]{carbonell1998use}
Jaime Carbonell and Jade Goldstein.
\newblock {The use of MMR, diversity-based reranking for reordering documents
  and producing summaries}.
\newblock In \emph{SIGIR}, 1998.

\bibitem[Chen et~al.(2018)Chen, Zhang, and Zhou]{chen2018dpp}
Laming Chen, Guoxin Zhang, and Eric Zhou.
\newblock {Fast greedy MAP inference for determinantal point process to improve
  recommendation diversity}.
\newblock In \emph{NeurIPS}, 2018.

\bibitem[Cully(2021)]{cully2021gecco}
Antoine Cully.
\newblock Multi-emitter {MAP-Elites}: Improving quality, diversity and data
  efficiency with heterogeneous sets of emitters.
\newblock In \emph{GECCO}, 2021.

\bibitem[Dai et~al.(2026)Dai, Meinardus, Regan, Tian, and Tang]{dai2026iclr}
Andrew Dai, Boris Meinardus, Ciaran Regan, Yingtao Tian, and Yujin Tang.
\newblock Discovering novel {LLM} experts via task-capability coevolution.
\newblock In \emph{ICLR}, 2026.

\bibitem[Dharna et~al.(2025)Dharna, Lu, and Clune]{dharna2025foundation}
Aaron Dharna, Cong Lu, and Jeff Clune.
\newblock Foundation model self-play: Open-ended strategy innovation via
  foundation models.
\newblock In \emph{RLC}, 2025.

\bibitem[Earle et~al.(2026)Earle, Arulkumaran, Kumar, Dai, Togelius, and
  Risi]{earle2026gecco}
Sam Earle, Kai Arulkumaran, Akarsh Kumar, Andrew Dai, Julian Togelius, and
  Sebastian Risi.
\newblock In search of the ingredients of open-endedness: Replicating
  {Picbreeder} with large vision-language models.
\newblock In \emph{GECCO}, 2026.

\bibitem[Gaier et~al.(2020)Gaier, Asteroth, and Mouret]{gaier2020gecco}
Adam Gaier, Alexander Asteroth, and Jean-Baptiste Mouret.
\newblock Discovering representations for black-box optimization.
\newblock In \emph{GECCO}, 2020.

\bibitem[Gallotta et~al.(2024)Gallotta, Arulkumaran, and Soros]{gallotta2024tg}
Roberto Gallotta, Kai Arulkumaran, and L.~B. Soros.
\newblock Preference-learning emitters for mixed-initiative quality-diversity
  algorithms.
\newblock \emph{IEEE Transactions on Games}, 16\penalty0 (2):\penalty0
  303--316, 2024.

\bibitem[Grillotti and Cully(2022)]{grillotti2022unsupervised}
Luca Grillotti and Antoine Cully.
\newblock Unsupervised behavior discovery with quality-diversity optimization.
\newblock \emph{IEEE Transactions on Evolutionary Computation}, 26\penalty0
  (6):\penalty0 1539--1552, 2022.

\bibitem[Guo et~al.(2024)Guo, Wang, Guo, Li, Song, Tan, Liu, Bian, and
  Yang]{guo2024iclr}
Qingyan Guo, Rui Wang, Junliang Guo, Bei Li, Kaitao Song, Xu~Tan, Guoqing Liu,
  Jiang Bian, and Yujiu Yang.
\newblock Connecting large language models with evolutionary algorithms yields
  powerful prompt optimizers.
\newblock In \emph{ICLR}, 2024.

\bibitem[Hu et~al.(2025)Hu, Lu, and Clune]{hu2025iclr}
Shengran Hu, Cong Lu, and Jeff Clune.
\newblock Automated design of agentic systems.
\newblock In \emph{ICLR}, 2025.

\bibitem[Hughes et~al.(2024)Hughes, Dennis, Parker-Holder, Behbahani,
  Mavalankar, Shi, Schaul, and Rockt{\"a}schel]{hughes2024icml}
Edward Hughes, Michael~D Dennis, Jack Parker-Holder, Feryal Behbahani, Aditi
  Mavalankar, Yuge Shi, Tom Schaul, and Tim Rockt{\"a}schel.
\newblock Position: Open-endedness is essential for artificial superhuman
  intelligence.
\newblock In \emph{ICML}, pages 20597--20616, 2024.

\bibitem[Jiang et~al.(2025)Jiang, Chai, Li, Liu, Fok, Dziri, Tsvetkov, Sap,
  Albalak, and Choi]{jiang2025artificial}
Liwei Jiang, Yuanjun Chai, Margaret Li, Mickel Liu, Raymond Fok, Nouha Dziri,
  Yulia Tsvetkov, Maarten Sap, Alon Albalak, and Yejin Choi.
\newblock Artificial hivemind: The open-ended homogeneity of language models
  (and beyond).
\newblock In \emph{NeurIPS Datasets and Benchmarks Track}, 2025.

\bibitem[Kaminskas and Bridge(2016)]{kaminskas2016diversity}
Marius Kaminskas and Derek Bridge.
\newblock {Diversity, serendipity, novelty, and coverage: A survey and
  empirical analysis of beyond-accuracy objectives in recommender systems}.
\newblock \emph{ACM Transactions on Interactive Intelligent Systems (TiiS)},
  7\penalty0 (1):\penalty0 1--42, 2016.

\bibitem[Kuroki et~al.(2025)Kuroki, Nakamura, Akiba, and Tang]{kuroki2024agent}
So~Kuroki, Taishi Nakamura, Takuya Akiba, and Yujin Tang.
\newblock Agent skill acquisition for large language models via {CycleQD}.
\newblock In \emph{ICLR}, 2025.

\bibitem[Lange et~al.(2026)Lange, Imajuku, and Cetin]{lange2026iclr}
Robert~Tjarko Lange, Yuki Imajuku, and Edoardo Cetin.
\newblock {ShinkaEvolve}: Towards open-ended and sample-efficient program
  evolution.
\newblock In \emph{ICLR}, 2026.

\bibitem[Lattimore and Szepesv{\'a}ri(2020)]{lattimore2020bandit}
Tor Lattimore and Csaba Szepesv{\'a}ri.
\newblock \emph{Bandit Algorithms}.
\newblock Cambridge University Press, 2020.

\bibitem[Lehman and Stanley(2011)]{lehman2011abandoning}
Joel Lehman and Kenneth~O Stanley.
\newblock Abandoning objectives: Evolution through the search for novelty
  alone.
\newblock \emph{Evolutionary computation}, 19\penalty0 (2):\penalty0 189--223,
  2011.

\bibitem[Lehman et~al.(2023)Lehman, Gordon, Jain, Ndousse, Yeh, and
  Stanley]{lehman2023evolution}
Joel Lehman, Jonathan Gordon, Shawn Jain, Kamal Ndousse, Cathy Yeh, and
  Kenneth~O Stanley.
\newblock Evolution through large models.
\newblock In \emph{Handbook of evolutionary machine learning}, pages 331--366.
  Springer, 2023.

\bibitem[Li et~al.(2010)Li, Chu, Langford, and Schapire]{li2010www}
Lihong Li, Wei Chu, John Langford, and Robert~E. Schapire.
\newblock A contextual-bandit approach to personalized news article
  recommendation.
\newblock In \emph{WWW}, 2010.

\bibitem[Lin et~al.(2026)Lin, Deldjoo, Dai, Bao, Ye, Nazary, Wang, Noia, Xu,
  and Chua]{lin2026arxiv}
Xinyu Lin, Yashar Deldjoo, Sunhao Dai, Honghui Bao, Xiaopeng Ye, Fatemeh
  Nazary, Wenjie Wang, Tommaso~Di Noia, Jun Xu, and Tat-Seng Chua.
\newblock Autonomous information seeking: A roadmap for agentic recommender
  systems.
\newblock \emph{arXiv preprint arXiv:2607.04433}, 2026.

\bibitem[Liu et~al.(2020)Liu, Xiao, Wu, Miao, Zhang, Zhao, and
  Tang]{liu2020aaai}
Yong Liu, Yingtai Xiao, Qiong Wu, Chunyan Miao, Juyong Zhang, Binqiang Zhao,
  and Haihong Tang.
\newblock Diversified interactive recommendation with implicit feedback.
\newblock In \emph{AAAI}, 2020.

\bibitem[Lu et~al.(2024)Lu, Holt, Fanconi, Chan, Foerster, van~der Schaar, and
  Lange]{lu2024discovering}
Chris Lu, Samuel Holt, Claudio Fanconi, Alex~J. Chan, Jakob Foerster, Mihaela
  van~der Schaar, and Robert~Tjarko Lange.
\newblock Discovering preference optimization algorithms with and for large
  language models.
\newblock In \emph{NeurIPS}, 2024.

\bibitem[Lu et~al.(2025)Lu, Hu, and Clune]{lu2025automated}
Cong Lu, Shengran Hu, and Jeff Clune.
\newblock Automated capability discovery via foundation model self-exploration.
\newblock \emph{arXiv preprint arXiv:2502.07577}, 2025.

\bibitem[McNee et~al.(2006)McNee, Riedl, and Konstan]{mcnee2006being}
Sean~M McNee, John Riedl, and Joseph~A Konstan.
\newblock {Being accurate is not enough: how accuracy metrics have hurt
  recommender systems}.
\newblock In \emph{CHI Extended Abstracts}, 2006.

\bibitem[Meyer and Corneil(2025)]{nemotron-personas-usa}
Yev Meyer and Dane Corneil.
\newblock {Nemotron-Personas-USA}: Synthetic personas aligned to real-world
  distributions, June 2025.
\newblock URL
  \url{https://huggingface.co/datasets/nvidia/Nemotron-Personas-USA}.

\bibitem[Mouret and Clune(2015)]{mouret2015illuminating}
Jean-Baptiste Mouret and Jeff Clune.
\newblock Illuminating search spaces by mapping elites.
\newblock \emph{arXiv preprint arXiv:1504.04909}, 2015.

\bibitem[Novikov et~al.(2025)Novikov, V{\~u}, Eisenberger, Dupont, Huang,
  Wagner, Shirobokov, Kozlovskii, Ruiz, Mehrabian,
  et~al.]{novikov2025alphaevolve}
Alexander Novikov, Ng{\^a}n V{\~u}, Marvin Eisenberger, Emilien Dupont, Po-Sen
  Huang, Adam~Zsolt Wagner, Sergey Shirobokov, Borislav Kozlovskii, Francisco
  J~R Ruiz, Abbas Mehrabian, et~al.
\newblock {AlphaEvolve: A coding agent for scientific and algorithmic
  discovery}.
\newblock \emph{arXiv preprint arXiv:2506.13131}, 2025.

\bibitem[{Paper Copilot (Jing Yang)}(2026)]{paperlists}
{Paper Copilot (Jing Yang)}.
\newblock {paperlists}: Processed / cleaned data for {Paper Copilot}.
\newblock \url{https://github.com/papercopilot/paperlists}, 2026.
\newblock GitHub repository.

\bibitem[Radlinski et~al.(2008)Radlinski, Kleinberg, and
  Joachims]{radlinski2008icml}
Filip Radlinski, Robert Kleinberg, and Thorsten Joachims.
\newblock Learning diverse rankings with multi-armed bandits.
\newblock In \emph{ICML}, 2008.

\bibitem[Risi et~al.(2025)Risi, Tang, Ha, and
  Miikkulainen]{risi-et-al:book2025}
Sebastian Risi, Yujin Tang, David Ha, and Risto Miikkulainen.
\newblock \emph{Neuroevolution: Harnessing Creativity in AI Agent Design}.
\newblock MIT Press, Cambridge, MA, 2025.

\bibitem[Romera-Paredes et~al.(2024)Romera-Paredes, Barekatain, Novikov, Balog,
  Kumar, Dupont, Ruiz, Ellenberg, Wang, Fawzi, et~al.]{romera2024funsearch}
Bernardino Romera-Paredes, Mohammadamin Barekatain, Alexander Novikov, Matej
  Balog, M~Pawan Kumar, Emilien Dupont, Francisco J~R Ruiz, Jordan~S Ellenberg,
  Pengming Wang, Omar Fawzi, et~al.
\newblock {Mathematical discoveries from program search with large language
  models}.
\newblock \emph{Nature}, 625\penalty0 (7995):\penalty0 468--475, 2024.

\bibitem[Samvelyan et~al.(2024)Samvelyan, Raparthy, Lupu, Hambro, Markosyan,
  Bhatt, Mao, Jiang, Parker-Holder, Foerster, Rockt{\"a}schel, and
  Raileanu]{samvelyan2024rainbow}
Mikayel Samvelyan, Sharath~Chandra Raparthy, Andrei Lupu, Eric Hambro, Aram~H
  Markosyan, Manish Bhatt, Yuning Mao, Minqi Jiang, Jack Parker-Holder, Jakob
  Foerster, Tim Rockt{\"a}schel, and Roberta Raileanu.
\newblock Rainbow teaming: Open-ended generation of diverse adversarial
  prompts.
\newblock In \emph{NeurIPS}, 2024.

\bibitem[Sfikas et~al.(2021)Sfikas, Liapis, and Yannakakis]{sfikas2021gecco}
Konstantinos Sfikas, Antonios Liapis, and Georgios~N. Yannakakis.
\newblock {Monte Carlo Elites}: Quality-diversity selection as a multi-armed
  bandit problem.
\newblock In \emph{GECCO}, 2021.

\bibitem[Shang et~al.(2025)Shang, Liu, Yan, Wu, Sheng, Yu, Jiang, Zhang, Xu,
  Wang, Zhang, and Li]{shang2025neurips}
Yu~Shang, Peijie Liu, Yuwei Yan, Zijing Wu, Leheng Sheng, Yuanqing Yu, Chumeng
  Jiang, An~Zhang, Fengli Xu, Yu~Wang, Min Zhang, and Yong Li.
\newblock {AgentRecBench}: Benchmarking {LLM} agent-based personalized
  recommender systems.
\newblock In \emph{NeurIPS Datasets and Benchmarks Track}, 2025.

\bibitem[Song et~al.(2022)Song, Sun, Lian, Huang, Li, Jin, and
  Xie]{song2022wsdm}
Yu~Song, Shuai Sun, Jianxun Lian, Hong Huang, Yu~Li, Hai Jin, and Xing Xie.
\newblock Show me the whole world: Towards entire item space exploration for
  interactive personalized recommendations.
\newblock In \emph{WSDM}, 2022.

\bibitem[Stanley and Lehman(2015)]{stanley2015book}
Kenneth~O. Stanley and Joel Lehman.
\newblock \emph{Why Greatness Cannot Be Planned: The Myth of the Objective}.
\newblock Springer, 2015.

\bibitem[Vargas and Castells(2011)]{vargas2011rank}
Sa{\'u}l Vargas and Pablo Castells.
\newblock {Rank and relevance in novelty and diversity metrics for recommender
  systems}.
\newblock In \emph{RecSys}, 2011.

\bibitem[Wan and McAuley(2018)]{wan2018goodreads}
Mengting Wan and Julian McAuley.
\newblock Item recommendation on monotonic behavior chains.
\newblock In \emph{RecSys}, 2018.

\bibitem[Wang et~al.(2026)Wang, Pi{\k{e}}kos, Nanbo, Laakom, Chen, Ostaszewski,
  Zhuge, and Schmidhuber]{wang2026iclr}
Wenyi Wang, Piotr Pi{\k{e}}kos, Li~Nanbo, Firas Laakom, Yimeng Chen, Mateusz
  Ostaszewski, Mingchen Zhuge, and J{\"u}rgen Schmidhuber.
\newblock {Huxley-G{\"o}del Machine}: Human-level coding agent development by
  an approximation of the optimal self-improving machine.
\newblock In \emph{ICLR}, 2026.

\bibitem[Wataoka et~al.(2024)Wataoka, Takahashi, and Ri]{wataoka2025neuripsws}
Koki Wataoka, Tsubasa Takahashi, and Ryokan Ri.
\newblock Self-preference bias in {LLM-as-a-Judge}.
\newblock In \emph{NeurIPS Safe Generative AI Workshop}, 2024.

\bibitem[Zelikman et~al.(2024)Zelikman, Lorch, Mackey, and
  Kalai]{zelikman2023stop}
Eric Zelikman, Eliana Lorch, Lester Mackey, and Adam~Tauman Kalai.
\newblock Self-taught optimizer ({STOP}): Recursively self-improving code
  generation.
\newblock In \emph{COLM}, 2024.

\bibitem[Zhang et~al.(2026{\natexlab{a}})Zhang, Hu, Lu, Lange, and
  Clune]{zhang2026iclr}
Jenny Zhang, Shengran Hu, Cong Lu, Robert Lange, and Jeff Clune.
\newblock {Darwin G{\"o}del Machine}: Open-ended evolution of self-improving
  agents.
\newblock In \emph{ICLR}, 2026{\natexlab{a}}.

\bibitem[Zhang et~al.(2026{\natexlab{b}})Zhang, Zhao, Yang, Foerster, Clune,
  Jiang, Devlin, and Shavrina]{zhang2026arxiv}
Jenny Zhang, Bingchen Zhao, Wannan Yang, Jakob Foerster, Jeff Clune, Minqi
  Jiang, Sam Devlin, and Tatiana Shavrina.
\newblock {HyperAgents}.
\newblock \emph{arXiv preprint arXiv:2603.19461}, 2026{\natexlab{b}}.

\bibitem[Zhang et~al.(2021)Zhang, Zhou, Li, and Gu]{zhang2021iclr}
Weitong Zhang, Dongruo Zhou, Lihong Li, and Quanquan Gu.
\newblock Neural {Thompson} sampling.
\newblock In \emph{ICLR}, 2021.

\bibitem[Zhang and Chen(2020)]{zhang2020explainable}
Yongfeng Zhang and Xu~Chen.
\newblock Explainable recommendation: A survey and new perspectives.
\newblock \emph{Foundations and Trends{\textregistered} in Information
  Retrieval}, 14\penalty0 (1):\penalty0 1--101, 2020.

\bibitem[Zhou et~al.(2020)Zhou, Li, and Gu]{zhou2020icml}
Dongruo Zhou, Lihong Li, and Quanquan Gu.
\newblock Neural contextual bandits with {UCB}-based exploration.
\newblock In \emph{ICML}, 2020.

\bibitem[Ziegler et~al.(2005)Ziegler, McNee, Konstan, and
  Lausen]{ziegler2005improving}
Cai-Nicolas Ziegler, Sean~M McNee, Joseph~A Konstan, and Georg Lausen.
\newblock {Improving recommendation lists through topic diversification}.
\newblock In \emph{WWW}, 2005.

\end{thebibliography}

\appendix

\section{Additional results: maps and curves over the run for all scenarios}
\label{app:results-table}
This appendix collects the qualitative maps for all eight scenarios (Figure~\ref{fig:maps-extra}) and the curves over the run, including semantic Diversity, for all eight scenarios (Figure~\ref{fig:qd-trajectory-extra}), with weaker-proposer (\texttt{claude-haiku-4-5}) counterparts in Figures~\ref{fig:maps-extra-haiku} and~\ref{fig:qd-trajectory-extra-haiku}.

\subsection{Qualitative maps: all scenarios}
\label{app:maps-extra}
\begin{figure*}[!ht]
    \centering
    \subfigure[Goodreads: series/YA reader (grounding)]{\includegraphics[width=0.49\textwidth]{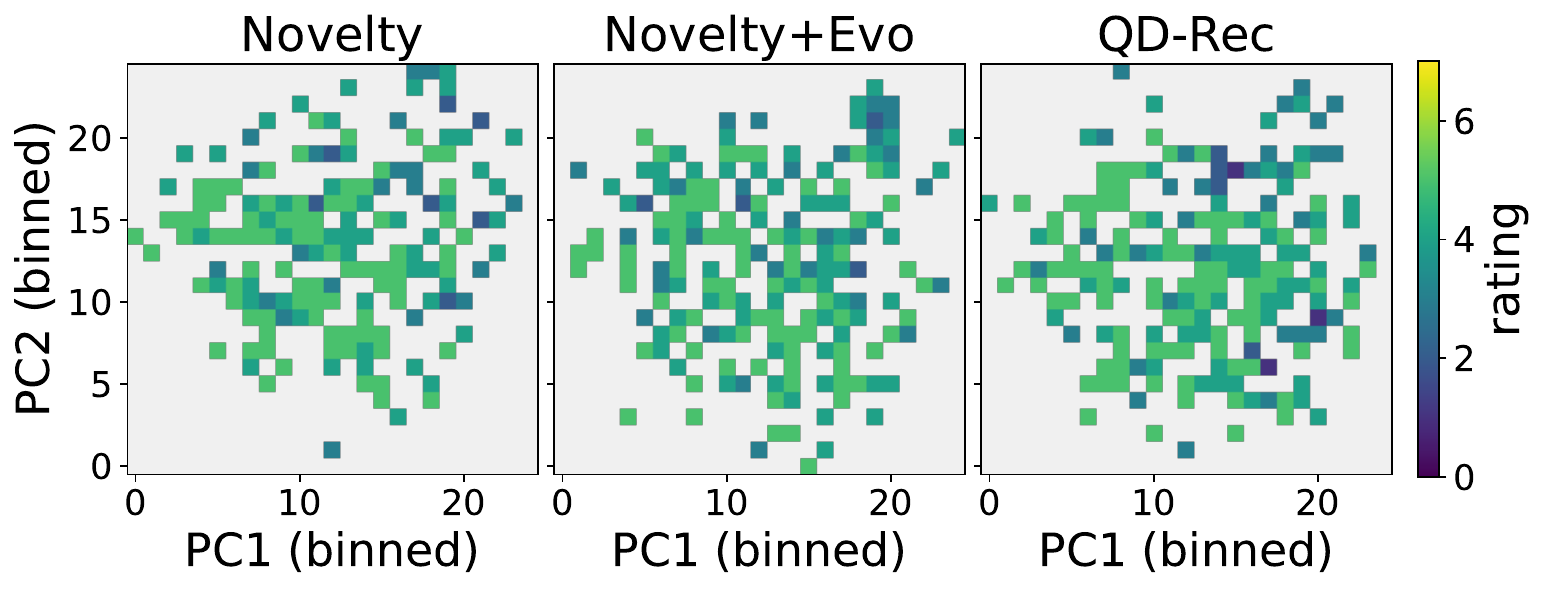}\label{fig:maps-syr}}
    \hfill
    \subfigure[Goodreads: picture-book reader]{\includegraphics[width=0.49\textwidth]{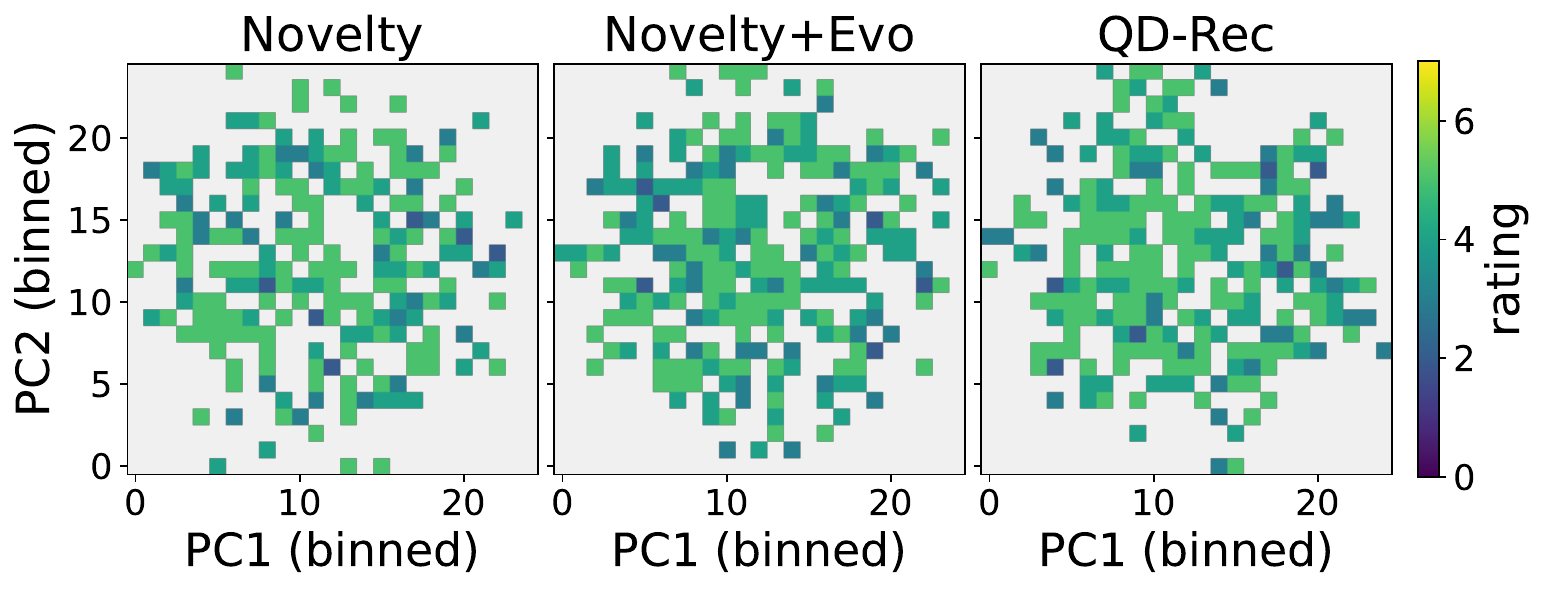}\label{fig:maps-pbr}}
    \vspace{-0.6em}

    \subfigure[M\&A: Walt Disney Co]{\includegraphics[width=0.49\textwidth]{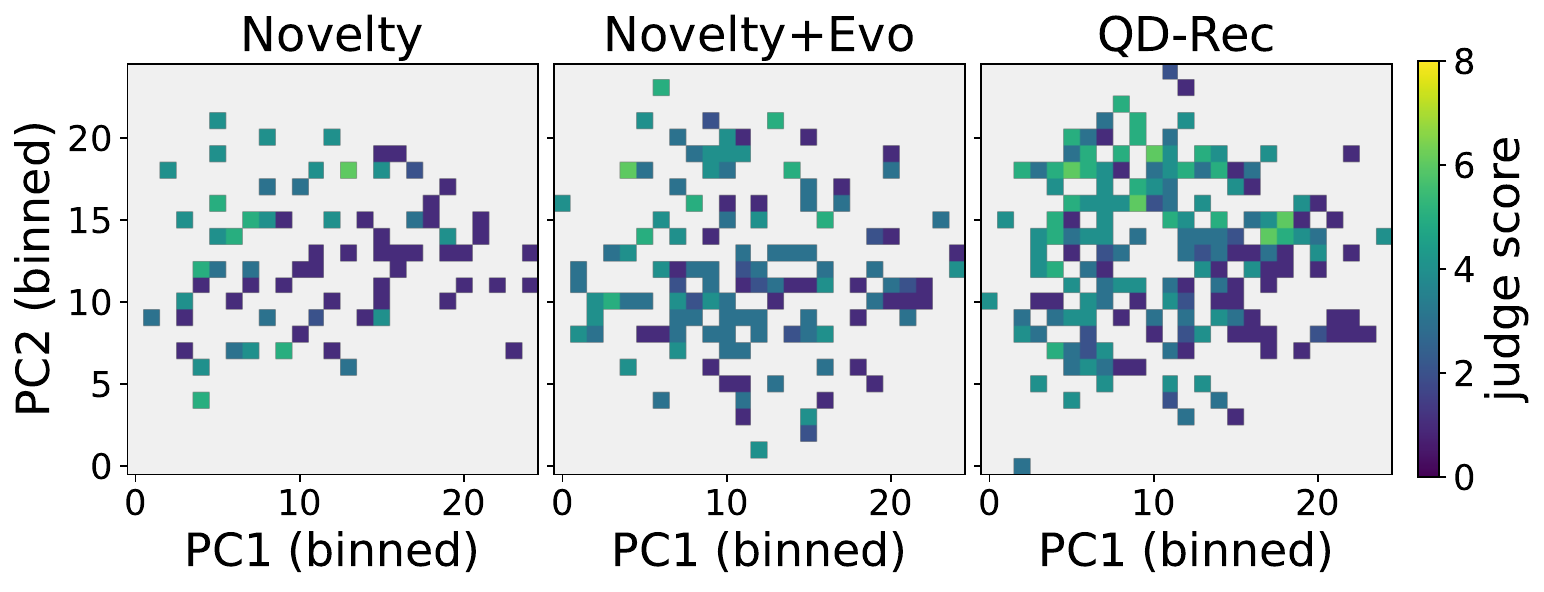}\label{fig:maps-dis-all3}}
    \hfill
    \subfigure[M\&A: JPMorgan Chase \& Co]{\includegraphics[width=0.49\textwidth]{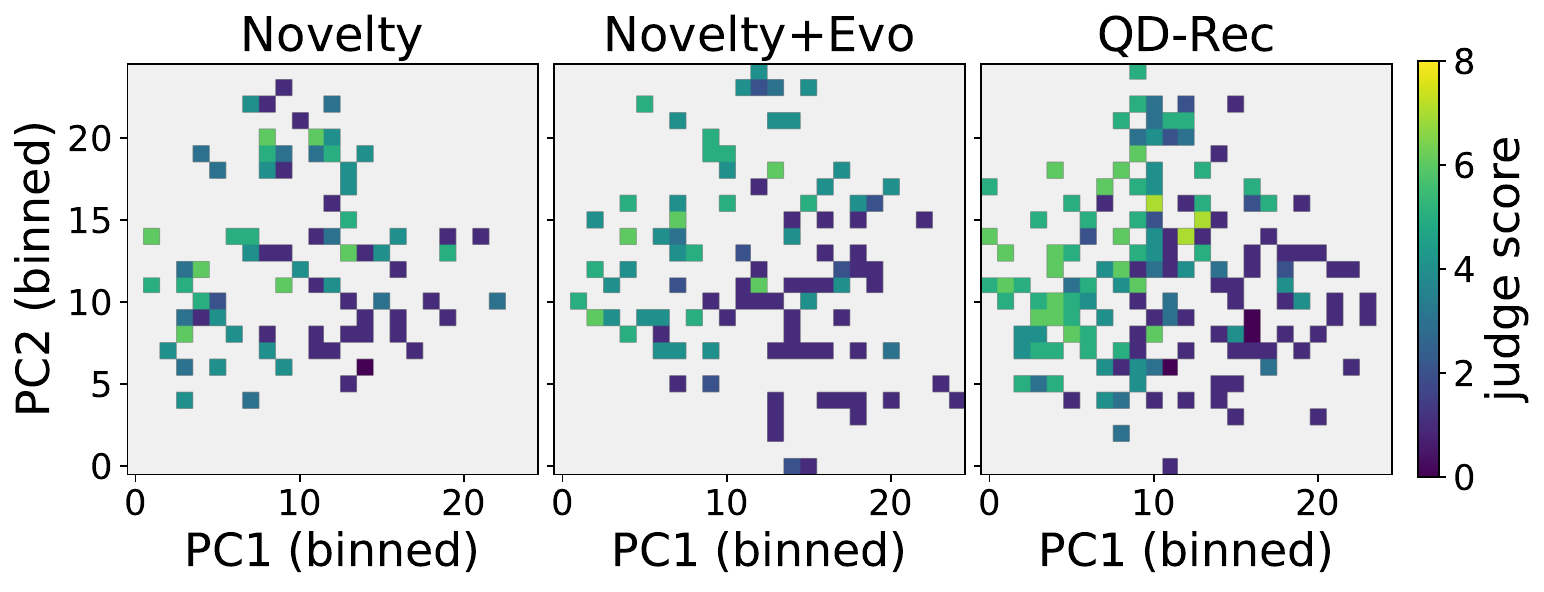}\label{fig:maps-jpm}}
    \vspace{-0.6em}

    \subfigure[ML Papers: OpenEndedness]{\includegraphics[width=0.49\textwidth]{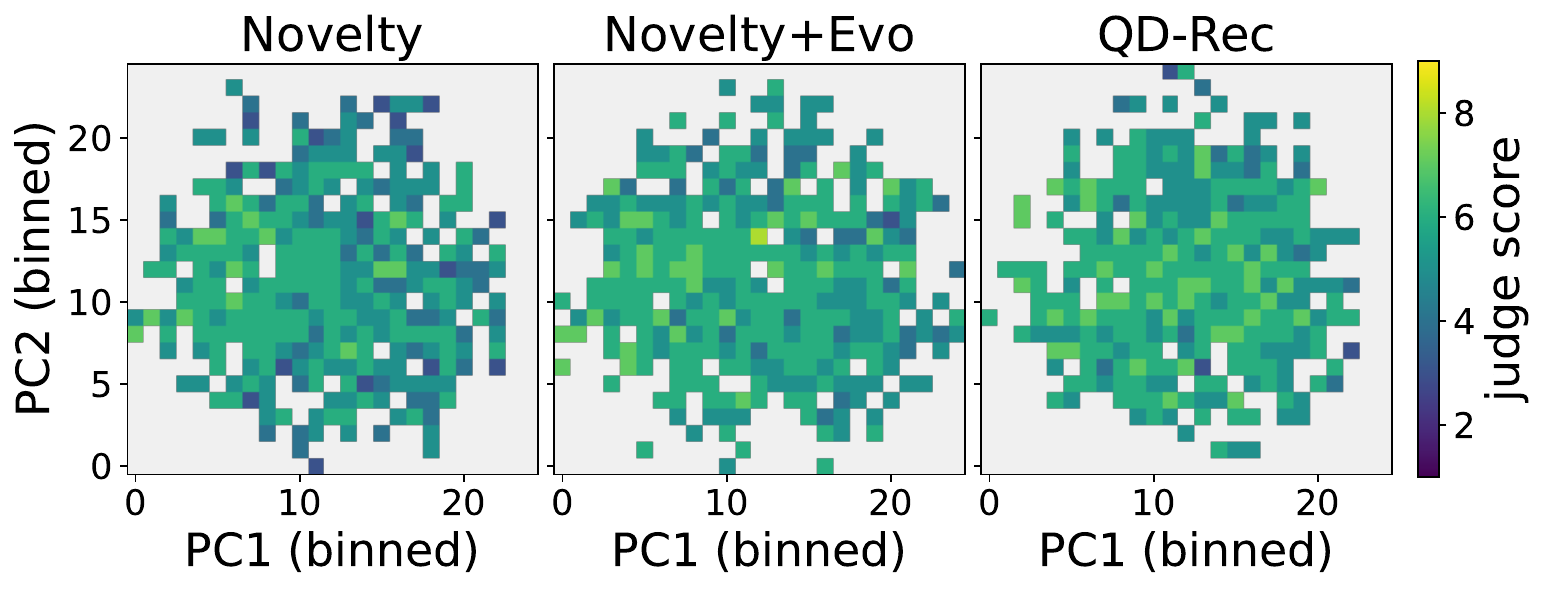}\label{fig:maps-oe-all3}}
    \hfill
    \subfigure[ML Papers: AI Startup]{\includegraphics[width=0.49\textwidth]{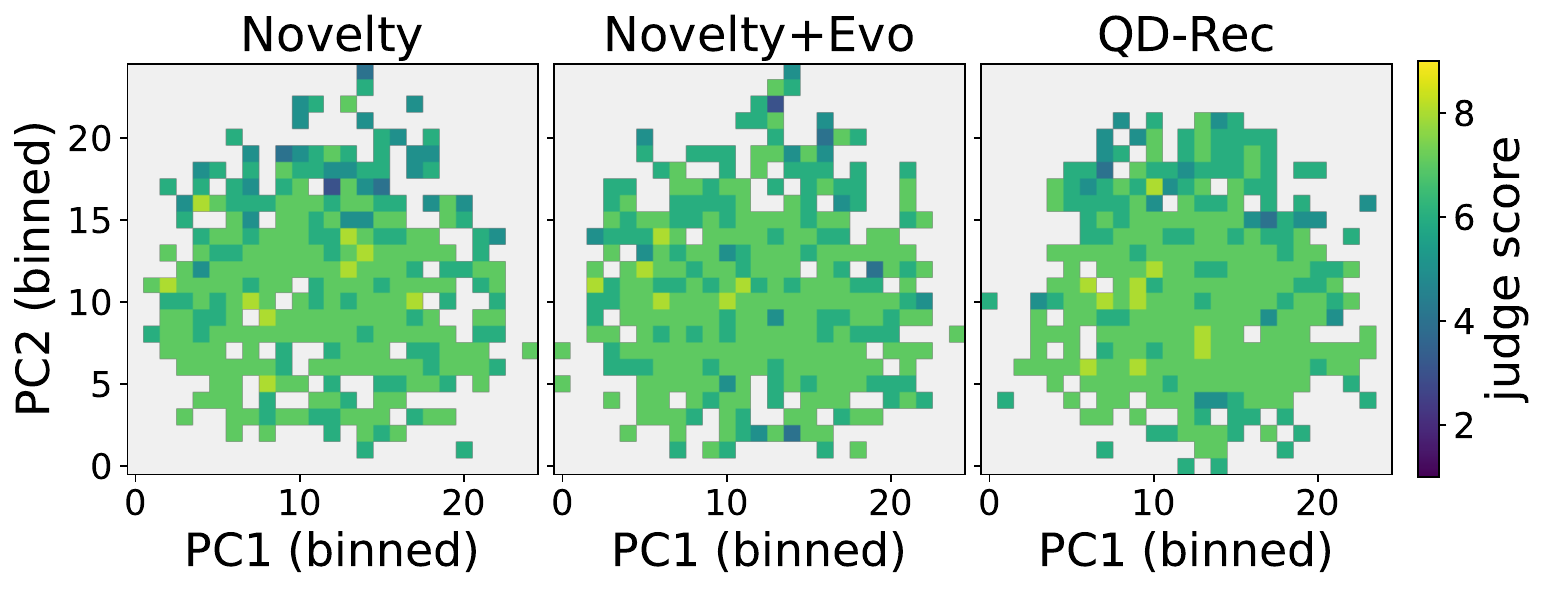}\label{fig:maps-startupai}}
    \vspace{-0.6em}

    \subfigure[Marketing: 3D printer]{\includegraphics[width=0.49\textwidth]{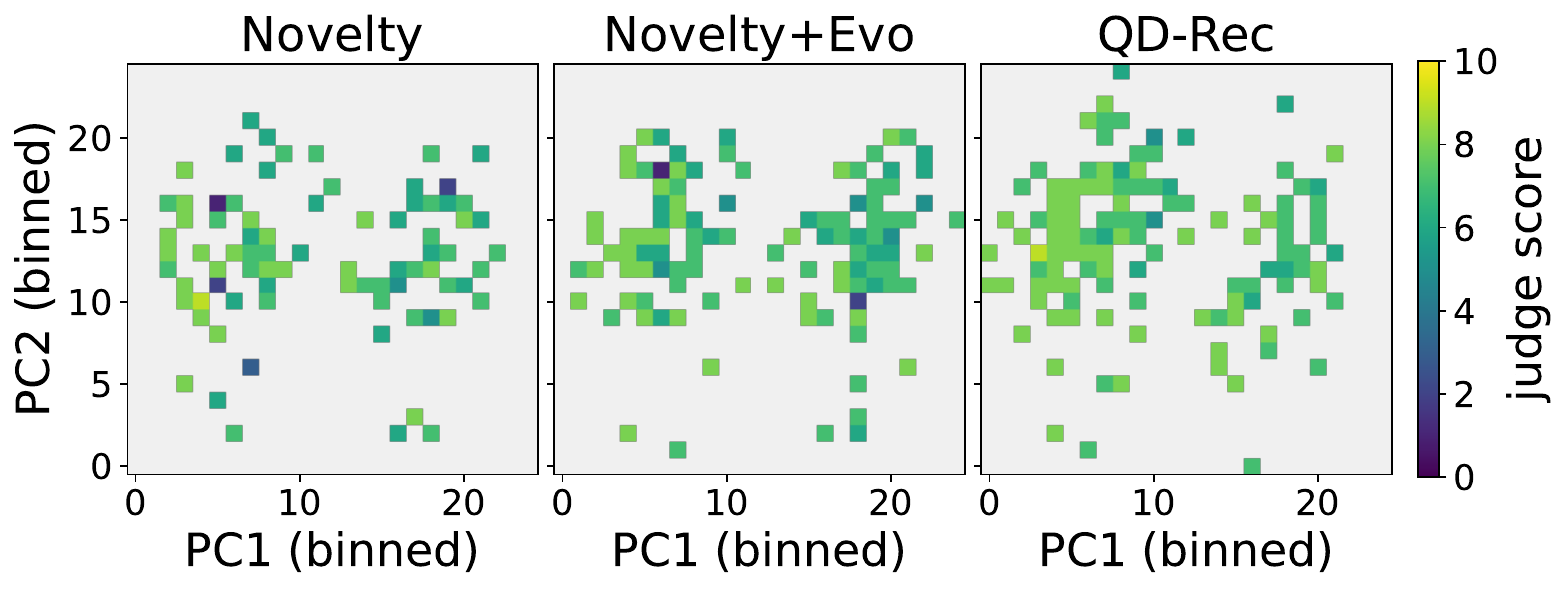}\label{fig:maps-printer3d}}
    \hfill
    \subfigure[Marketing: Upskilling]{\includegraphics[width=0.49\textwidth]{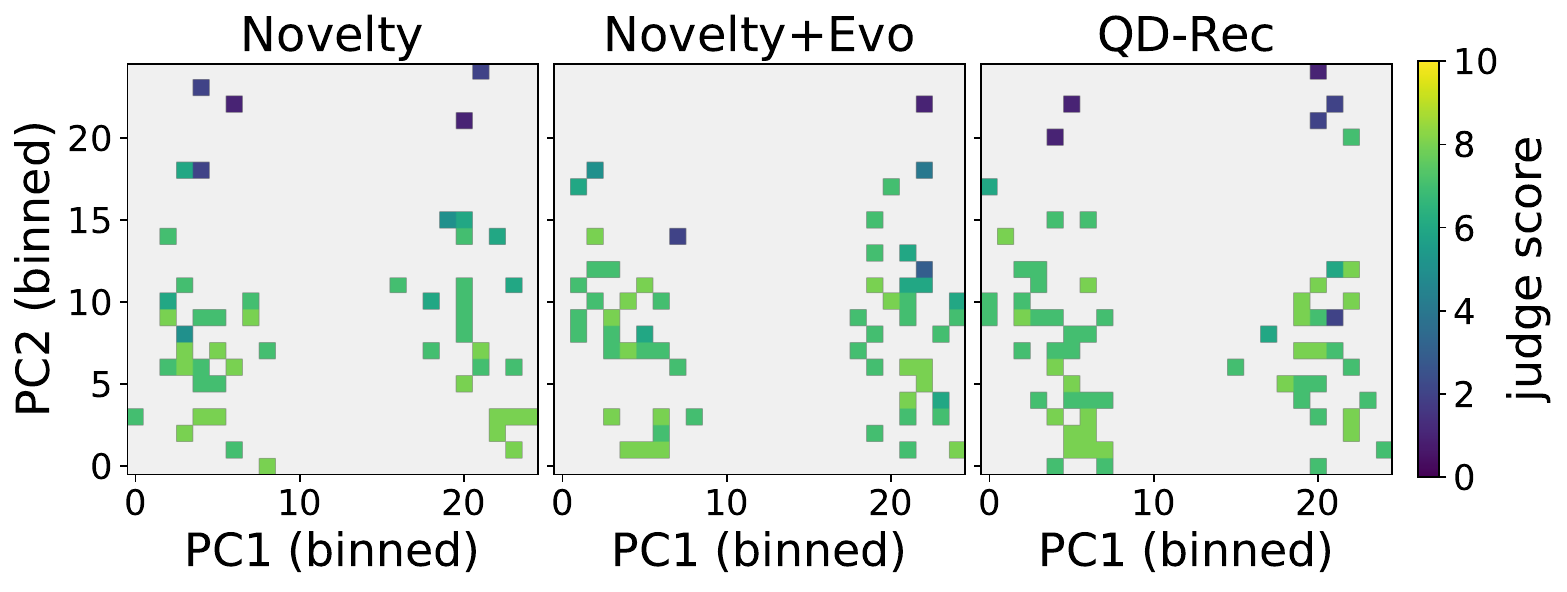}\label{fig:maps-skillforge}}

    \caption{Qualitative maps for all eight scenarios (both profiles of every domain), complementing Figure~\ref{fig:main-map}. Same construction: $25{\times}25$ MAP-Elites-style heatmaps (best quality per PCA-binned cell), here showing all three variants (Novelty, Novelty+Evo, QD-Rec); color ranges are auto-scaled per domain. All panels use the seed-$0$ \texttt{claude-sonnet-5}-proposer runs. Goodreads uses human star ratings rather than an LLM judge.}
    \label{fig:maps-extra}
\end{figure*}

\begin{figure*}[!ht]
    \centering
    \subfigure[Goodreads: series/YA reader (grounding)]{\includegraphics[width=0.49\textwidth]{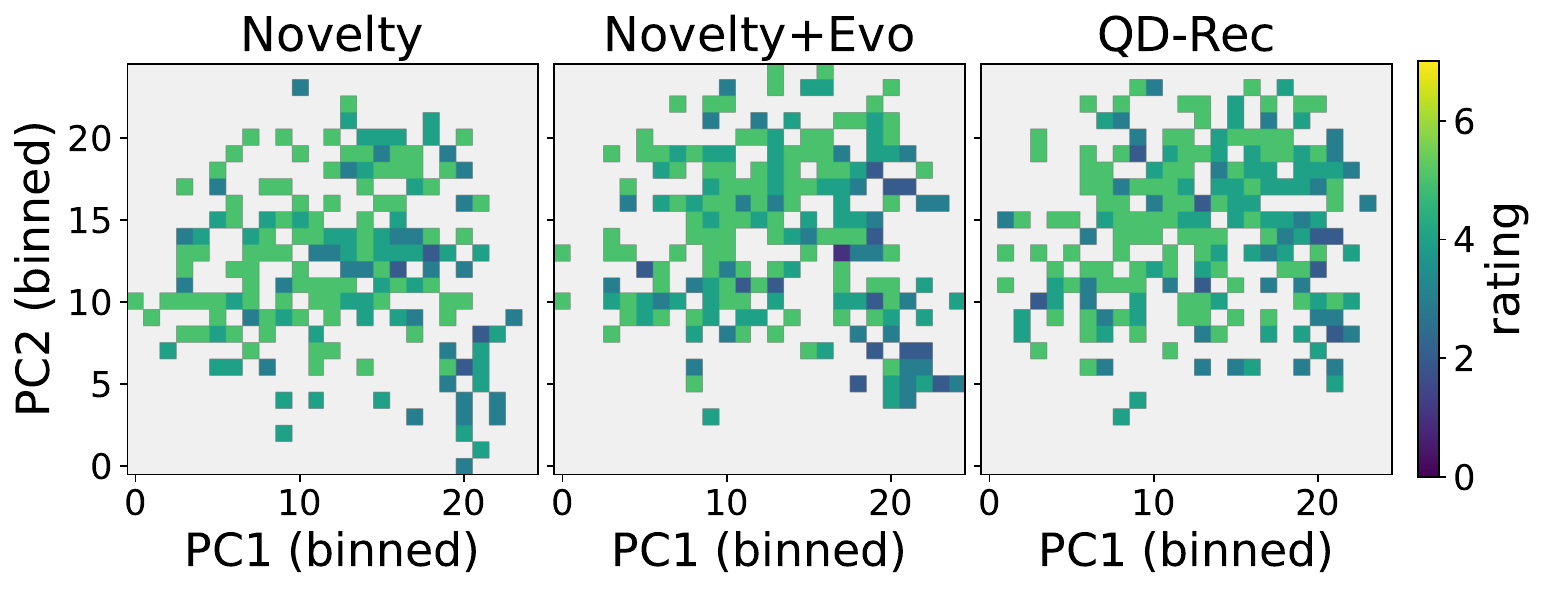}}
    \hfill
    \subfigure[Goodreads: picture-book reader]{\includegraphics[width=0.49\textwidth]{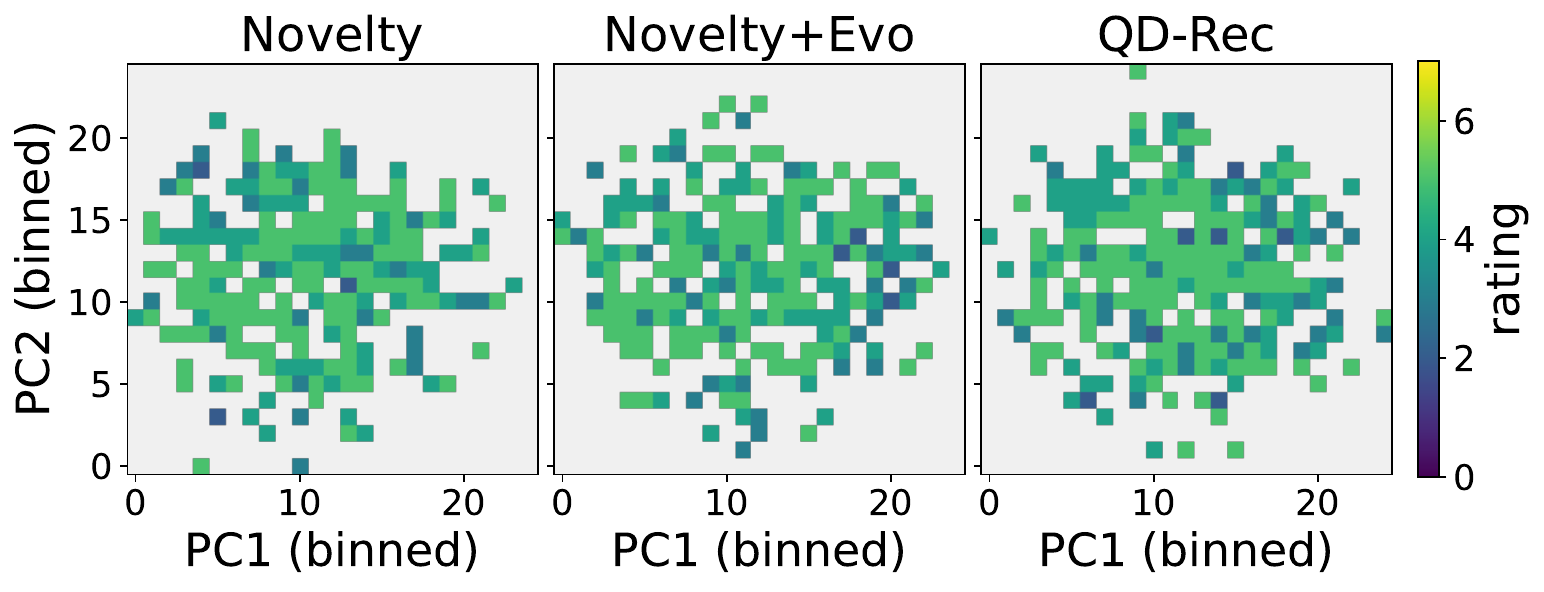}}
    \vspace{-0.6em}

    \subfigure[M\&A: Walt Disney Co]{\includegraphics[width=0.49\textwidth]{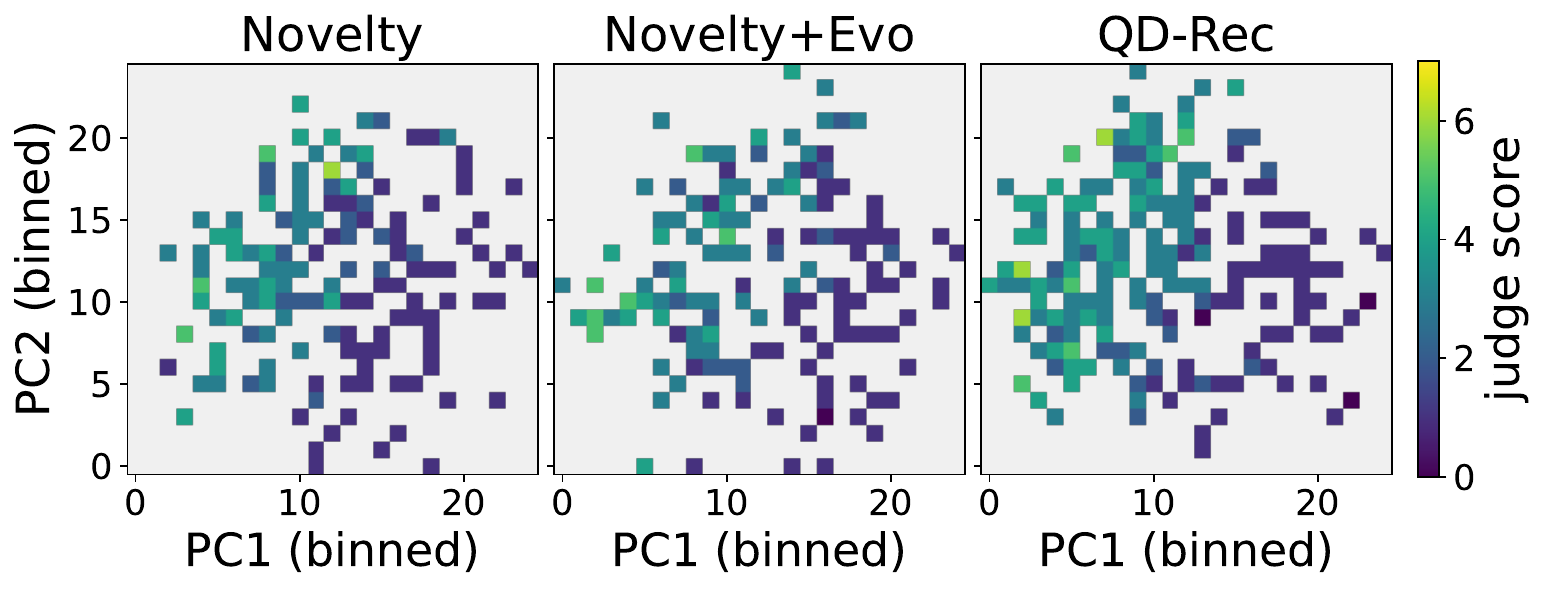}}
    \hfill
    \subfigure[M\&A: JPMorgan Chase \& Co]{\includegraphics[width=0.49\textwidth]{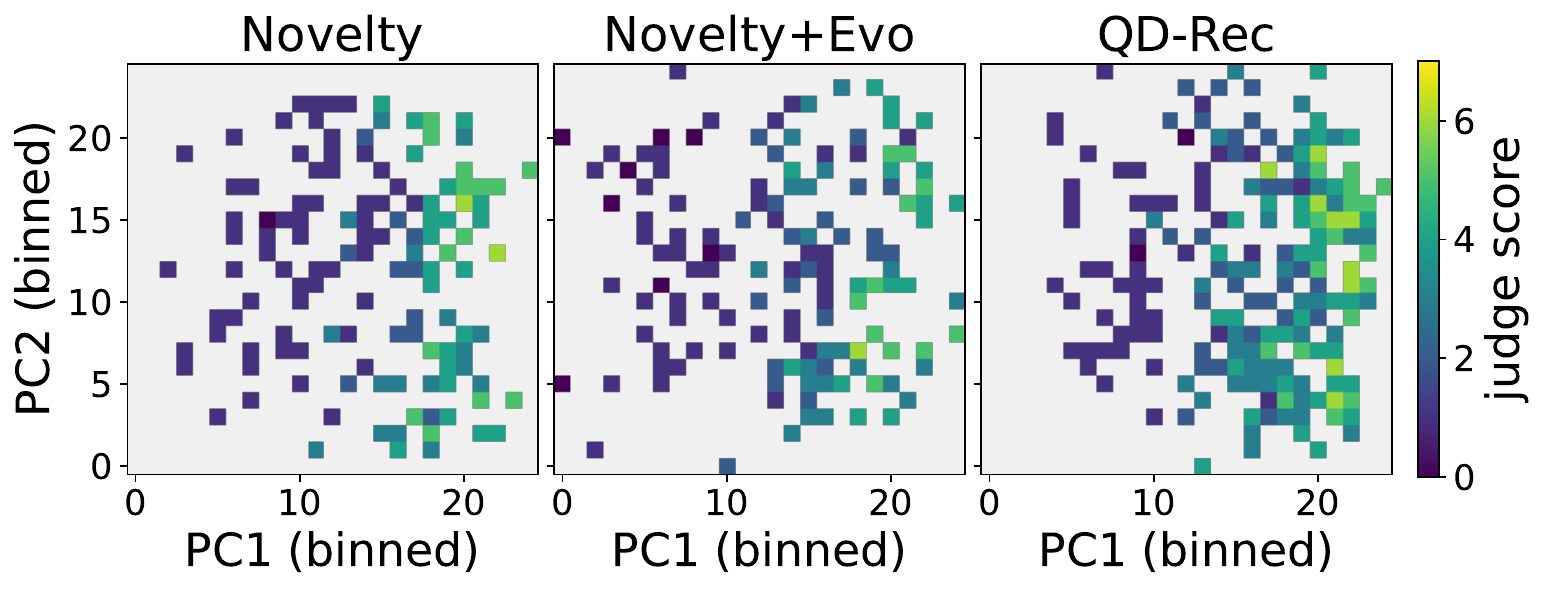}}
    \vspace{-0.6em}

    \subfigure[ML Papers: OpenEndedness]{\includegraphics[width=0.49\textwidth]{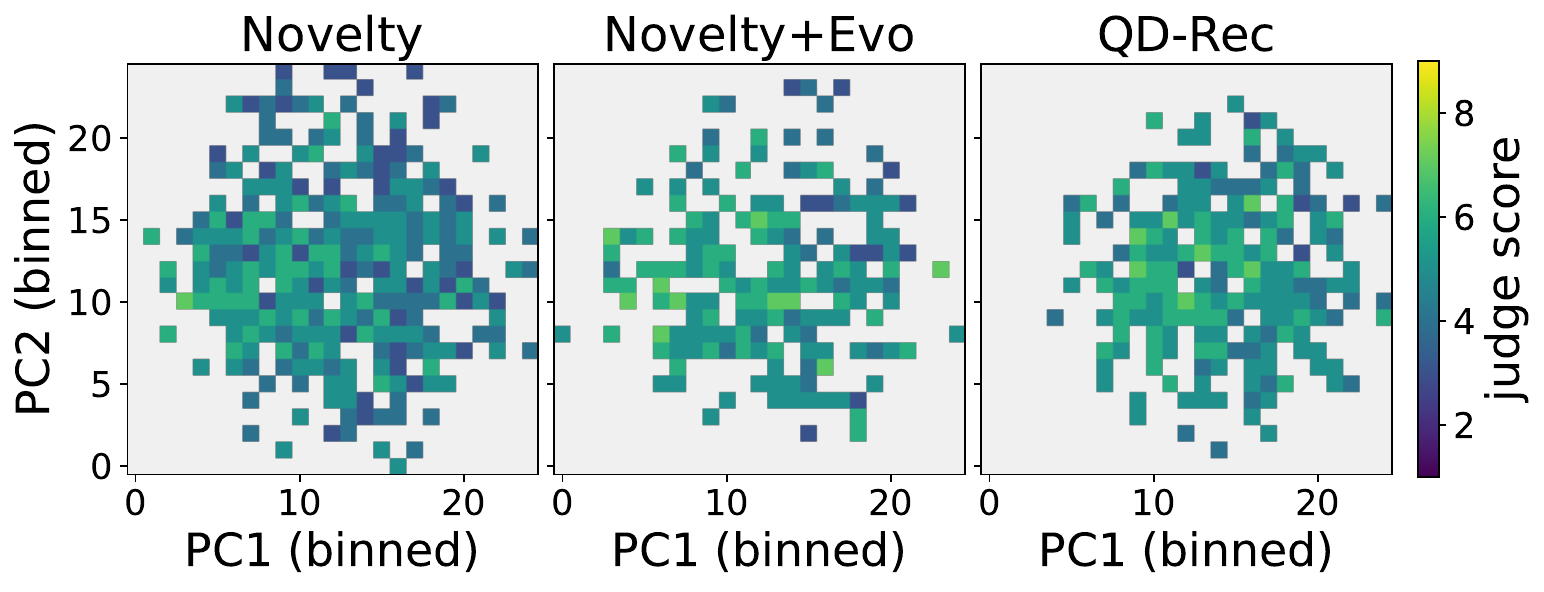}}
    \hfill
    \subfigure[ML Papers: AI Startup]{\includegraphics[width=0.49\textwidth]{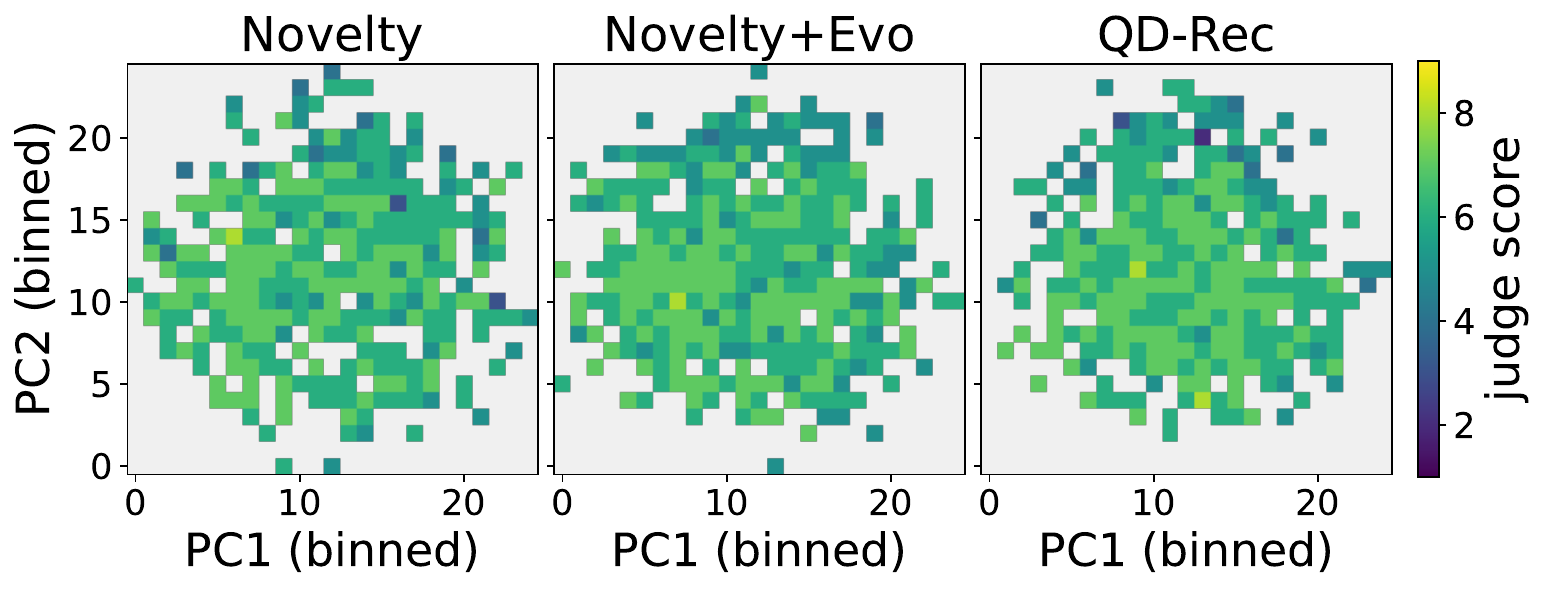}}
    \vspace{-0.6em}

    \subfigure[Marketing: 3D printer]{\includegraphics[width=0.49\textwidth]{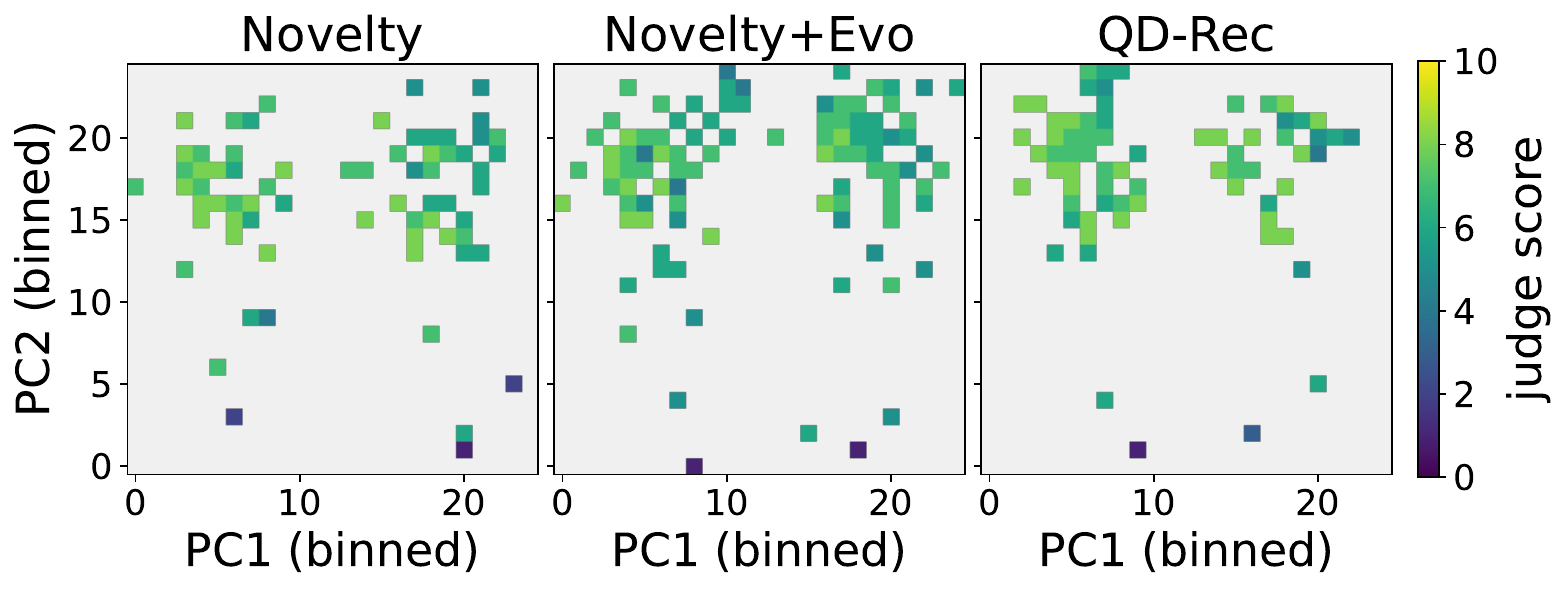}}
    \hfill
    \subfigure[Marketing: Upskilling]{\includegraphics[width=0.49\textwidth]{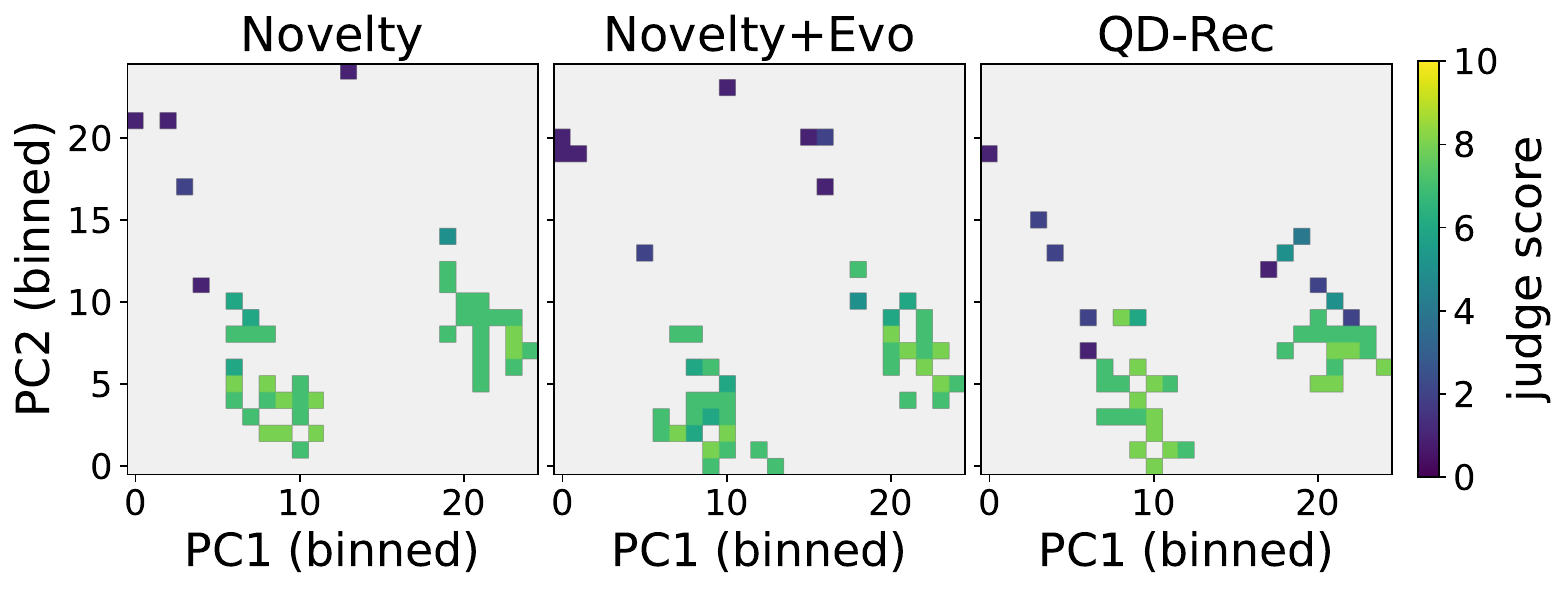}}

    \caption{Qualitative maps for all eight scenarios under the weaker proposer \texttt{claude-haiku-4-5}, complementing Figure~\ref{fig:maps-extra}. Same construction: seed-$0$ runs, all three variants, $25{\times}25$ best-quality-per-cell heatmaps, with color ranges auto-scaled per domain within this proposer's runs.}
    \label{fig:maps-extra-haiku}
\end{figure*}

Figure~\ref{fig:maps-extra} shows the MAP-Elites-style heatmaps of Figure~\ref{fig:main-map} for \emph{all} eight scenarios (both profiles of every domain), each with all three variants. Figure~\ref{fig:maps-extra-haiku} repeats the construction for the weaker-proposer ablation (\texttt{claude-haiku-4-5}).

\subsection{Metrics over the run: all scenarios}
\label{app:trajectory-extra}
\begin{figure*}[!ht]
\centering
\includegraphics[width=\linewidth]{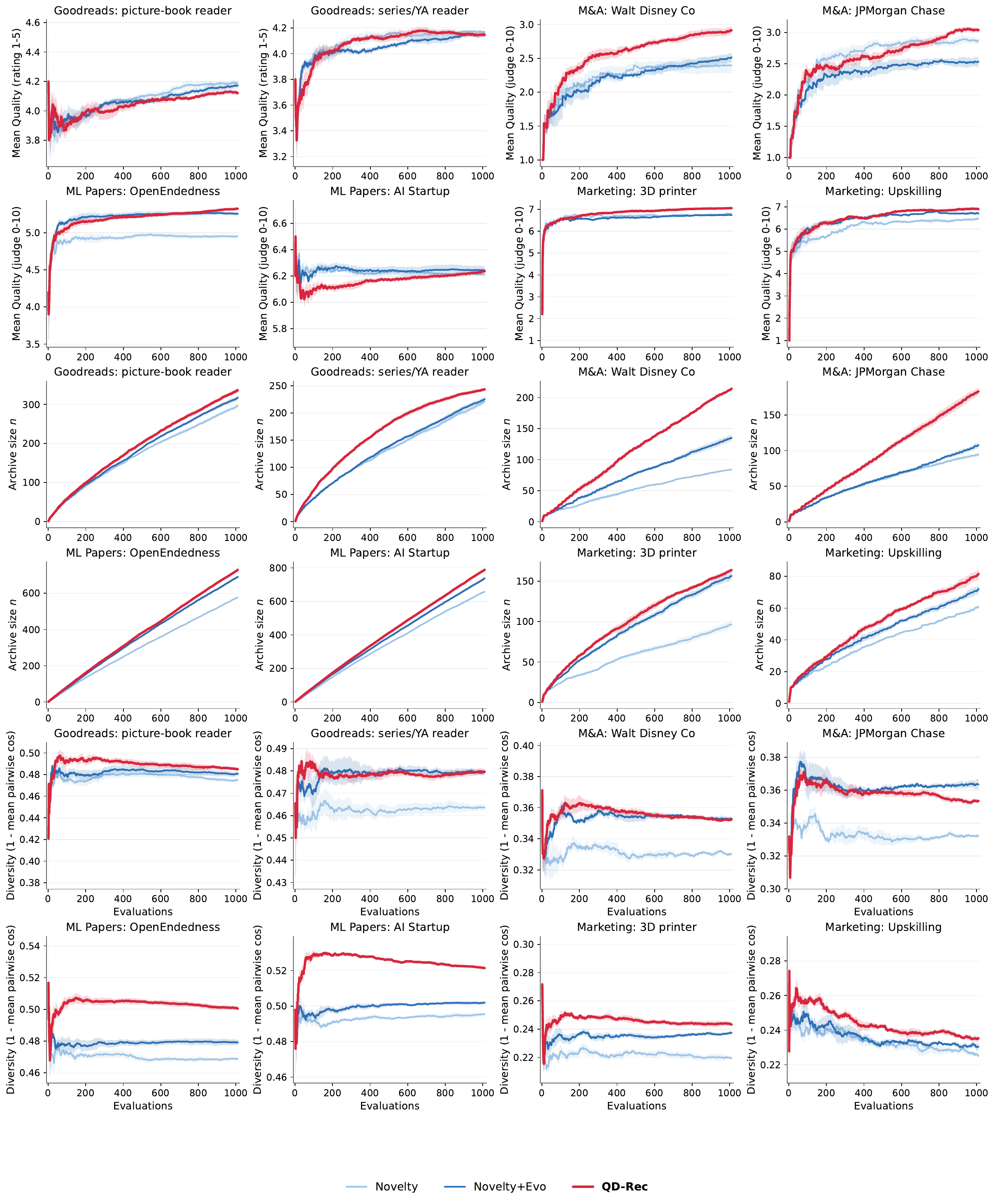}
\caption{Metrics over the run for all eight scenarios (same construction and \texttt{claude-sonnet-5}-proposer runs as Figure~\ref{fig:main}). Rows 1--2 show mean Quality for the eight scenarios, rows 3--4 the corresponding archive size $n$, and rows 5--6 semantic Diversity; panel titles identify the scenario. Goodreads Quality is near-flat because it is the reader's fixed $1$--$5$ rating.}
\label{fig:qd-trajectory-extra}
\end{figure*}

\begin{figure*}[!ht]
\centering
\includegraphics[width=\linewidth]{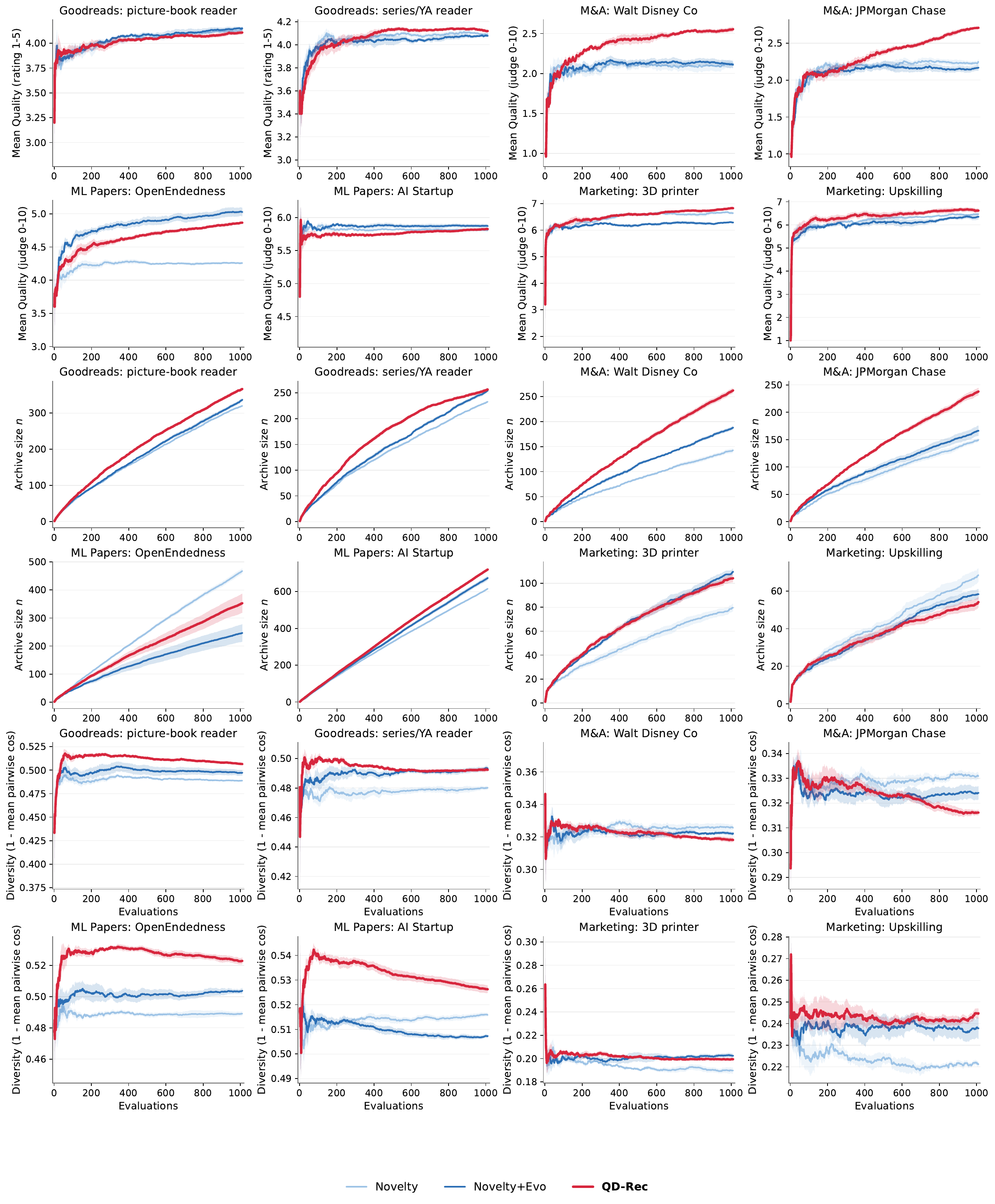}
\caption{Metrics over the run for all eight scenarios under the weaker proposer (\texttt{claude-haiku-4-5}; same construction as Figure~\ref{fig:qd-trajectory-extra}). Rows 1--2 show mean Quality, rows 3--4 archive size $n$, and rows 5--6 semantic Diversity; panel titles identify the scenario.}
\label{fig:qd-trajectory-extra-haiku}
\end{figure*}

Figure~\ref{fig:qd-trajectory-extra} plots mean Quality, archive size $n$, and semantic Diversity against the number of evaluations for all eight scenarios. Figure~\ref{fig:qd-trajectory-extra-haiku} shows the same curves under the weaker proposer \texttt{claude-haiku-4-5}.

Unlike Quality, Diversity changes little after the first ${\sim}100$ evaluations, while archive size $n$ keeps growing. A likely reason is that admission is decided by the LLM novelty check against nearest neighbors, not by an embedding-distance threshold. New entries then add distinct hypotheses within the region the archive already spans, which raises $n$ but not the mean pairwise distance.

\section{No-learning baselines: relevance-MMR and a quality-weighted DPP}
\label{app:mmr}
To ask whether the online learning and archive machinery earns its cost, we compare against two training-free re-rankers on Goodreads, where Quality is a real human star rating that neither the learning methods nor the baselines can game. \emph{Relevance-MMR} \citep{carbonell1998use} greedily selects items by $\gamma\cos(\text{item},\pi)-(1-\gamma)\max_{j\in S}\cos(\text{item},j)$ over the same content embeddings, where $S$ is the set of items selected so far. \emph{Quality-weighted DPP} \citep{chen2018dpp} is the principled counterpart: greedy MAP inference for a determinantal point process maximizing $\gamma\sum_{i\in S} \mathrm{rel}_i + (1-\gamma)\log\det(L_S)$. Here $\mathrm{rel}_i=\cos(\text{item}_i,\pi)$ is the relevance of item $i$, the same relevance MMR uses. The matrix $L_S$ holds the pairwise cosine similarities of the items in $S$, so $\log\det(L_S)$ measures the diversity of the set as its log-volume. In both baselines, the weight $\gamma\in[0,1]$ trades relevance against diversity, and $\gamma{=}1$ recovers plain top-$K$ retrieval exactly.

Neither baseline uses an archive, novelty filter, or feedback, and no LLM touches anything that could affect the headline result: selection is pure geometry over precomputed content embeddings, with relevance $\cos(\text{item},\pi)$ rather than an LLM score (scoring the full pool with an LLM is precisely the cost our setting excludes), and Quality is the human star rating, fixed the moment an item is selected. The only LLM calls are measurement plumbing: after selection, each chosen item receives a zero-shot memo (proposer \texttt{claude-sonnet-5}) and the same joint item+memo embedding the QD entries carry, so that Diversity and Coverage are computed in the same embedding space as Table~\ref{tab:exp2-main} and every row emits the same item--hypothesis artifact. These memos cannot influence which items are selected or their ratings.

For each reader, the Coverage grid is fitted jointly to all QD, MMR, and DPP rows using full SVD and the same 1\% padding convention. This larger reference pool differs from Table~\ref{tab:exp2-main}, so absolute Coverage need not match across tables. The comparison is not a controlled ablation of item selection, however: the QD variants observe the rating of every item they evaluate (up to $1{,}000$ per run) and use it for replacement and, in QD-Rec, for the bandit, whereas the baselines see none. They are no-feedback geometric references, and the gap measures what rating feedback and the archive machinery buy together. Every run is matched in archive size $n$ to the QD-Rec archive of its reader; selection is deterministic given the embeddings, so rows are single runs (Table~\ref{tab:mmr}).

No setting of either baseline reaches any QD variant on Quality. The baselines approach QD-Rec's Diversity only at their most diversity-favoring settings, and a few settings exceed its Coverage. The DPP gives almost the same results as MMR, so the gap is not specific to MMR's heuristic.

\begin{table}[t]
  \centering
  \footnotesize
  \setlength{\tabcolsep}{5pt}
  \begin{tabular}{l cccc}
  \toprule
  Method & Qual. & Div. & Cov. & $n$ \\
  \midrule
  \multicolumn{5}{l}{\emph{Goodreads: picture-book reader}} \\
  Novelty                & \textbf{4.22} & .474 & .285 & 281 \\
  Novelty+Evo            & 4.13 & .477 & .318 & 322 \\
  QD-Rec                 & 4.12 & .484 & .326 & \textbf{344} \\
  MMR $\gamma{=}1.0$     & 3.69 & .408 & .312 & \textbf{344} \\
  MMR $\gamma{=}0.7$     & 3.70 & .420 & .326 & \textbf{344} \\
  MMR $\gamma{=}0.5$     & 3.69 & .442 & .328 & \textbf{344} \\
  MMR $\gamma{=}0.3$     & 3.68 & .478 & .318 & \textbf{344} \\
  DPP $\gamma{=}1.0$     & 3.69 & .408 & .323 & \textbf{344} \\
  DPP $\gamma{=}0.7$     & 3.71 & .453 & .314 & \textbf{344} \\
  DPP $\gamma{=}0.5$     & 3.71 & .477 & \textbf{.330} & \textbf{344} \\
  DPP $\gamma{=}0.3$     & 3.68 & \textbf{.484} & .328 & \textbf{344} \\
  \addlinespace
  \multicolumn{5}{l}{\emph{Goodreads: series/YA reader}} \\
  Novelty                & \textbf{4.18} & .467 & .234 & 206 \\
  Novelty+Evo            & 4.16 & \textbf{.481} & .246 & 231 \\
  QD-Rec                 & 4.12 & .481 & .248 & \textbf{234} \\
  MMR $\gamma{=}1.0$     & 3.31 & .401 & .229 & \textbf{234} \\
  MMR $\gamma{=}0.7$     & 3.48 & .420 & \textbf{.259} & \textbf{234} \\
  MMR $\gamma{=}0.5$     & 3.61 & .428 & .250 & \textbf{234} \\
  MMR $\gamma{=}0.3$     & 3.68 & .443 & .230 & \textbf{234} \\
  DPP $\gamma{=}1.0$     & 3.31 & .397 & .234 & \textbf{234} \\
  DPP $\gamma{=}0.7$     & 3.59 & .433 & .248 & \textbf{234} \\
  DPP $\gamma{=}0.5$     & 3.68 & .451 & .248 & \textbf{234} \\
  DPP $\gamma{=}0.3$     & 3.67 & .454 & .242 & \textbf{234} \\
  \bottomrule
  \end{tabular}
  \caption{Learning vs.\ no-learning on the human-rated domain. The three QD variants vs.\ the training-free relevance-MMR and quality-weighted DPP ($\gamma$ sweep) re-rankers on Goodreads (real $1$--$5$ star ratings). Every baseline run selects as many items as the QD-Rec archive of its reader holds and writes a zero-shot memo for each (proposer \texttt{claude-sonnet-5}); QD rows are the seed-$0$ runs of the same generation (\texttt{claude-sonnet-5} proposer). The baselines see no ratings, so they are no-feedback geometric references rather than a controlled ablation of item selection. Bold marks the best per metric within each reader block.}
  \label{tab:mmr}
\end{table}

\section{Dataset details}
\label{app:datasets}
This appendix gives the full dataset construction for the four domains of Section~\ref{sec:domains} (sources, filtering, pool sizes, and the specific profiles used).

\paragraph{Goodreads (real human-feedback grounding).}
We use the Children subset of the UCSD Goodreads Book Graph \citep{wan2018goodreads}. For each reader, the candidate pool is their own rated books (minus the few used to construct the profile), so every item carries a ground-truth rating. We select two readers by a fixed procedure, set before any experiment. A reader qualifies if they meet six thresholds: at least $200$ rated books, at least $500$ written reviews, a rating standard deviation of at least $0.9$, at least $70\%$ of shelved books marked as read, at least $80\%$ of sampled reviews in English, and an entropy of at least $2.0$ nats in their books' distribution over shelves. Of the $542{,}145$ readers, $8$ qualify. We take the qualifying reader with the most ratings, and then the one with the most ratings whose most common shelf differs from the first reader's. This gives a picture-book reader ($1{,}761$ books) and a series/young-adult reader ($1{,}117$ books). Goodreads uses the reader's $1$--$5$ rating in place of the LLM judge, so the bandit reward is $q(x)/5$. In the local comparison, a new book replaces its nearest neighbor only if its rating is strictly higher; ties keep the incumbent. Books are selected without replacement in all three variants, including seed books and discarded proposals, because a fixed rating means a new memo for the same book would get the same score.

\paragraph{M\&A target discovery.}
We construct country-specific company universes from LSEG Workspace data that combine (i) tabular financial snapshot features (as of 2024-08-31) and (ii) textual business descriptions. We filter to a target country, retain feature columns with at least 0.80 fill rate, and merge description fields into a single company profile. This yields universes of 6{,}332 USA companies. The profile $\pi$ is an acquirer selected from the same universe; in our runs we use two USA acquirers, Walt Disney Co and JPMorgan Chase \& Co. For QD-Rec, each candidate company has a precomputed embedding of its business description, used only for bandit item selection. The proposer, the judge and the archive embedding use the full company profile, which also includes the location, the industry classification and the financial features.

\paragraph{Marketing.}
We use a subset of the NVIDIA Nemotron-Personas-USA dataset, a large-scale \emph{synthetic customer-profile} corpus aligned to real-world US demographic distributions \citep{nemotron-personas-usa}. We use this synthetic dataset as a proxy in lieu of sensitive real-world customer data. Each customer profile includes demographic attributes (e.g., age, sex, marital status, education, occupation, city/state) and multiple profile facets (professional, sports, arts, travel, culinary). To control for regional heterogeneity we restrict to a single state (California) and sample 10{,}000 customer profiles for experiments. We use two distinct product profiles (a desktop 3D printer and an online upskilling subscription), spanning different purchase contexts (a one-time tool purchase vs.\ career reskilling). The seed archive is initialized from sampled customer profiles, and the three variants iteratively generate and evaluate targeting hypotheses for additional profiles.

\paragraph{Paper-abstract exploration.}
We use a fixed collection of paper metadata (title and abstract only) drawn from three recent machine-learning conferences: ICLR 2026, ICML 2026, and NeurIPS 2025. We filter to accepted statuses (Oral, Spotlight, and Poster, plus ICLR's conditional accepts) and retain entries with non-empty identifier, title, and abstract, yielding 17{,}511 papers (5{,}358 / 6{,}341 / 5{,}812 for ICLR / ICML / NeurIPS respectively). The metadata is obtained from the \texttt{papercopilot/paperlists} repository \citep{paperlists}. We precompute candidate embeddings from title+abstract for bandit item selection. Archive embeddings are computed from the concatenated title, abstract, and memo (\texttt{item\_hypothesis} mode) and are used for novelty-neighbor retrieval, semantic Diversity, and Coverage. The three variants generate memos for selected papers and apply the same local novelty-first archive update procedure.

\section{Prompt Templates}
\label{app:prompts}
This appendix lists the prompts used in our experiments. The three variants use the same generation and judging prompts. They differ only in how the next item is selected and whether the generator is shown archive exemplars. For each domain we show the generator prompt, the judge prompt (0--10 score), the novelty-check prompt, and the pairwise-duel prompt, which runs only when an entry is not novel. Goodreads has only the generator and novelty-check prompts, because it uses human ratings for Quality and replacement. Placeholders such as \texttt{\{acquirer\_profile\}} are filled in at run time. A novelty-check or duel response that cannot be parsed never adds an entry to the archive.

\subsection{Example Profiles Used in Experiments}
\label{app:example-profiles}

\begin{lstlisting}[style=prompt,caption={Researcher profile used in ML Papers experiments: OpenEndedness Researcher.},label={lst:profile-openendedness}]
An ML researcher working on open-ended and quality-diversity methods. Their through-line is divergent search, reaching ambitious capabilities through diverse intermediate stepping stones rather than direct climbs toward a fixed objective, but in practice they bring a handful of separable questions to a paper. Sometimes a paper is interesting for how it measures novelty or behavioral diversity without hand-designed descriptors, through learned representations and embeddings that define what a good behavior space even is; sometimes for its archive and selection mechanics, such as what to retain and how local competition and illumination shape search; sometimes for its operators, such as mutation, crossover, or using large models as variation operators; sometimes for auto-curricula and co-evolution, where agents and the environments that challenge them generate their own next problems; sometimes for what it says about exploration and the deception of objectives, including intrinsic motivation, when direct optimization stalls, and what transfers from reinforcement-learning exploration; and sometimes for recursive self-improvement, where systems rewrite their own code, operators, or objectives so that gains compound, as in self-referential and self-modifying agents. They have a skeptical streak toward pure scaling and benchmark-chasing, and they actively look to import techniques from adjacent areas into open-ended search. They do not expect every paper at a major ML conference to be about open-endedness; they look for the specific angle (a method, a measurement, a failure mode, an assumption) that connects to their interest.
\end{lstlisting}

\begin{lstlisting}[style=prompt,caption={Researcher profile used in ML Papers experiments: AI Startup Researcher.},label={lst:profile-startupai}]
An applied ML researcher at an early-stage AI startup who reads research through the lens of practical leverage. They are not only asking, "Could this become a product?" but also, "Could this make a product cheaper, safer, faster, more reliable, more defensible, easier to evaluate, or easier to specialize?" Their interest is in papers that could plausibly affect a 3-to-12-month startup roadmap, whether by enabling a new product surface, improving unit economics, reducing inference cost, increasing reliability, strengthening safety guarantees, opening a vertical market, or creating a data or evaluation moat. When reading papers from major ML conferences in particular, they do not expect every paper to be immediately productizable. They recognize that much of this work is theoretical, exploratory, or several steps removed from company use, so they look for the subset of work that hints at future engineering leverage, changes an important assumption, improves the efficiency/safety/reliability envelope, or suggests a direction that could compound into practical advantage over time.
\end{lstlisting}

\begin{lstlisting}[style=prompt,caption={Acquirer profile used in M\&A experiments: Walt Disney Co.},label={lst:mna-acquirer-disney}]
Company: Walt Disney Co
Location: Burbank, Americas (United States of America)

Classification:
  - Sector: Consumer Cyclicals / Cyclical Consumer Services
  - Industry: Media & Publishing / Broadcasting
  - Activity: Broadcasting (NEC)

Business Description:
The Walt Disney Company is a diversified worldwide entertainment company. The Company's segments include Entertainment, Sports and Experiences. The Entertainment segment generally encompasses its non-sports focused global film and episodic content production and distribution activities. The lines of business within the Entertainment segment along with their business activities include Linear Networks, Direct-to-Consumer, and Content Sales/Licensing. The Sports segment encompasses its sports-focused global television and direct-to-consumer (DTC) video streaming content production and distribution activities. The lines of business within the Sports segment include ESPN and Star. The Experiences segment includes Parks and Experiences and Consumer Products. Parks and Experiences consists of Walt Disney World Resort in Florida, Disneyland Resort in California, Disney Cruise Line, and others. Consumer Products includes licensing of its trade names, characters, visual, literary and other IP.

Financial Profile (Snapshot: 2024-08-31, Local Currency):
  - Enterprise Value: 210223027404.4 (Current)
  - Historical EV (1y/2y/3y/4y ago): 198446140333.96 / 257149691645.81 / 387047387650.601 / 296707312395.75
  - Performance: 0.562294155788656% YTD, 4.84709850044702% 2-Year
\end{lstlisting}

\begin{lstlisting}[style=prompt,caption={Acquirer profile used in M\&A experiments: JPMorgan Chase \& Co.},label={lst:mna-acquirer-jpm}]
Company: JPMorgan Chase & Co
Location: New York City, Americas (United States of America)

Classification:
  - Sector: Financials / Banking & Investment Services
  - Industry: Banking Services / Banks
  - Activity: Banks (NEC)

Business Description:
JPMorgan Chase & Co. is a financial holding company. The Company is engaged in investment banking, financial services for consumers and small businesses, commercial banking, financial transaction processing and asset management. The Company operates through three segments: Consumer & Community Banking (CCB), Commercial & Investment Bank (CIB), and Asset & Wealth Management (AWM). Its CCB segment offers products and services to consumers and small businesses through bank branches, ATMs, digital and telephone banking. Its CIB segment consists of banking and payments and markets and securities services, and offers a suite of investment banking, lending, payments, market-making, financing, custody and securities products and services to a global base of corporate and institutional clients. AWM segment offers investment and wealth management solutions. It offers multi-asset investment management solutions, retirement products and services, brokerage, custody, estate planning, and others.

Financial Profile (Snapshot: 2024-08-31, Local Currency):
  - Enterprise Value: 962614030629.599 (Current)
  - Historical EV (1y/2y/3y/4y ago): 619830439781.86 / 289049648861.661 / 420986173300.6 / 442185113328.18
  - Performance: 34.4997336773436% YTD, 75.6934494012104% 2-Year
\end{lstlisting}

\begin{lstlisting}[style=prompt,caption={Product profile used in Marketing experiments: 3D printer.},label={lst:product-printer3d-en}]
LAYERCRAFT is a fictional desktop 3D-printer brand in the $500-$900 price range. It is aimed at people who already make, fix, or teach things and want to do so with their own parts and designs, and it positions itself as "the printer that just prints": the machine arrives assembled and calibrated, and the first print starts within fifteen minutes of unboxing.

The lineup comprises three models. "Layercraft Studio" is a compact, fully enclosed printer, quiet enough for an apartment or a classroom. "Layercraft Workshop" has a larger build volume and prints tougher engineering materials for functional parts. "Layercraft Color" holds four spools and switches between them automatically for multi-color prints.

Every model has automatic bed leveling, a companion app with a library of ready-to-print designs, and a 2-year warranty with US-based support. A design marketplace lets owners sell or share their own models, and educators get free lesson plans and a classroom management mode. Filament is sold in recyclable spools at $20-$30 per kilogram.

Owners use it for different things: replacing a broken part on an appliance or a car instead of ordering one; prototyping a product for a small business or a side venture; STEM projects with children or students; props, cosplay and tabletop-gaming pieces; custom organizers, tools and gifts. It is not for everyone: it needs a permanent desk-sized spot, a few evenings to learn basic 3D design and slicing, filament at $20-$30 per kilogram, and patience, since a typical print takes several hours.

LAYERCRAFT frames a 3D printer as a tool for people who make things, like a sewing machine or a table saw: something that pays for itself in the things it makes and the things it fixes. By taking the setup and the failed prints out of the experience, it aims to make owning one ordinary in that kind of household.
\end{lstlisting}

\begin{lstlisting}[style=prompt,caption={Product profile used in Marketing experiments: Upskilling.},label={lst:product-skillforge-en}]
SkillForge is a fictional online learning platform offered at $89-$249 per month. Centered on business skills, technology, and creative fields, it provides content aimed at practical, job-ready skill acquisition. Its defining feature is that it is designed not as passive video-watching but around the goal of "being able to use what you learn at work."

Courses are organized into three learning tracks. The "Business Track" delivers skills that make you immediately useful in white-collar roles -- data analysis, project management, financial modeling, presentation. The "Tech Track" offers hands-on courses from the basics to applications of Python, SQL, cloud, and AI, understandable even for non-engineers. The "Creative Track" teaches expressive skills usable for side work and career changes, such as UI/UX design, video editing, and writing.

Each course is taught by a working expert and includes a weekly live Q&A session, a Slack community of fellow learners, and assignment submissions with feedback modeled on real work. Progress is visualized on a dashboard, and completion certificates can be posted on LinkedIn or a portfolio. A corporate plan adds team-level enrollment management and progress reporting.

The intended users are diverse: late-20s-to-30s professionals aiming to upskill in their current jobs, women considering reskilling as they plan a return from childcare, seniors preparing for a second career after retirement, and employees with a side-gig mindset who want a source of income beyond their main job.

With the concept of "making re-learning unremarkable," SkillForge is designed so that even busy working adults can steadily build skills with 3-4 hours of study a week. With no enrollment fee and flexible, cancel-anytime plans, it lowers the barrier to staying with learning and aims to support the career autonomy of working people in the US.
\end{lstlisting}

\subsection{M\&A target discovery}

\begin{lstlisting}[style=prompt,caption={M\&A few-shot hypothesis generation (system).},label={lst:mna-fewshot-gen}]
You are an M&A strategist who generates acquisition hypotheses.
\end{lstlisting}

\begin{lstlisting}[style=prompt,caption={M\&A few-shot hypothesis generation (user).},label={lst:mna-fewshot-gen-user}]
## Acquiring Company:
{acquirer_profile}

---

## Existing Acquisition Theses (for inspiration only):

{existing_theses}

---

## Target Company to Evaluate:
{target_profile}

---

Generate an acquisition hypothesis explaining why the acquirer should consider acquiring the target company.

Focus on:
- Strategic fit with the acquirer's existing business
- Specific synergies (revenue growth, cost savings, capability acquisition)
- Market positioning and competitive advantages
- Financial attractiveness (valuation, growth trajectory, deal sizing relative to acquirer)
- Timing considerations (is the target's current valuation/performance favorable?)

Be concrete and specific to the actual companies involved. Reference financial metrics when relevant (e.g., enterprise value, growth rates, returns).

Make the strongest plausible case for why the acquirer should acquire this company; recommend against it only if the strategic fit is genuinely poor.

Use the existing theses as inspiration. You may:
- **Mutate**: Take a strategic angle from ONE thesis and adapt it to this new target
- **Crossover**: Combine elements from TWO OR MORE theses into a novel rationale
- **Explore**: Propose a completely different strategic direction not yet covered by any thesis

Also report which operator you used and the thesis number(s) you built on, using the
"Thesis N" numbers shown above (leave the source list empty for Explore).

Do not copy phrases verbatim from the existing theses.

Provide your response in this EXACT JSON format:

{
  "operation": "mutate" | "crossover" | "explore",
  "source_examples": [thesis numbers you built on, e.g. [2] for mutate or [1, 3] for crossover; [] for explore],
  "evolution_note": "ONE sentence on how the example thesis/theses shaped this hypothesis, e.g. which angle came from which Thesis N. This is the ONLY field that may reference the examples; leave empty for explore.",
  "hypothesis": "2-6 sentences explaining WHY acquiring this company would benefit the acquirer. Be specific about strategic fit, synergies, and value creation potential. Write it as a STANDALONE rationale, in the same style as one written with no examples in front of you: do NOT reference, number, or allude to the example theses, and do NOT narrate the operator or how it was derived (put any such mention only in evolution_note)."
}
\end{lstlisting}

\begin{lstlisting}[style=prompt,caption={M\&A pointwise judge (system).},label={lst:mna-pointwise-judge}]
You are a critical M&A advisor scoring acquisition hypotheses.
\end{lstlisting}

\begin{lstlisting}[style=prompt,caption={M\&A pointwise judge (user).},label={lst:mna-pointwise-judge-user}]
Evaluate the following acquisition hypothesis.

## Acquiring Company:
{acquirer_profile}

## Target Company:
{target_profile}

## Acquisition Hypothesis:
{hypothesis}

---

Score how compelling and plausible the strategic case for the acquisition is, considering:
- Strategic fit with the acquirer's existing business
- Synergy potential (revenue and cost synergies)
- Feasibility (is this acquisition realistic and executable?)
- Risk factors (integration challenges, market risks, regulatory concerns)
- Value creation potential

A hypothesis that argues against the acquisition, concludes the target is a poor fit, or offers only a hollow/implausible rationale is a weak recommendation and must score low, even if its reasoning is accurate. Reserve high scores for genuinely compelling, specific, plausible theses.

Provide your evaluation in this EXACT JSON format:

{
  "reasoning": "Few sentences summarizing your overall assessment",
  "score": <integer from 0-10>
}

Scoring guide (strength of the recommendation to acquire):
- 0-2: No real case (argues against, no strategic fit, or hollow/implausible rationale)
- 3-4: Weak case (strained or generic rationale; major concerns)
- 5-6: Moderate case (some plausible merit but notable gaps)
- 7-8: Strong case (specific, plausible fit with clear value creation)
- 9-10: Excellent case (highly compelling, specific, plausible thesis)
\end{lstlisting}

\begin{lstlisting}[style=prompt,caption={M\&A novelty check (system).},label={lst:mna-novelty}]
You are a senior M&A advisor who assesses the novelty of acquisition theses.
\end{lstlisting}

\begin{lstlisting}[style=prompt,caption={M\&A novelty check (user).},label={lst:mna-novelty-user}]
## Acquiring Company:
{acquirer_profile}

---

## Existing Acquisition Theses:

{existing_text}

---

## New Thesis to Assess:

{new_text}

---

Assess whether the NEW thesis above represents a meaningfully different strategic direction or value-creation mechanism compared to the existing ones:
- Consider target industry, strategic rationale (vertical/horizontal integration, capability acquisition, market access), and synergy type (revenue, cost, technology, talent).
- Evaluate strictly: minor variations or superficially different targets with the same thesis are NOT novel. Only truly distinct strategic directions count.

Respond in the following JSON format:

{
  "novelty_reasoning": "1-2 sentences explaining whether the new thesis is a genuinely novel strategic direction compared to the existing ones.",
  "is_novel": true or false
}
\end{lstlisting}

\begin{lstlisting}[style=prompt,caption={M\&A blind pairwise duel (system).},label={lst:mna-duel}]
You are a senior M&A advisor who compares acquisition theses and decides which is stronger.
\end{lstlisting}

\begin{lstlisting}[style=prompt,caption={M\&A blind pairwise duel (user).},label={lst:mna-duel-user}]
## Acquiring Company:
{acquirer_profile}

---

## Acquisition Theses to Compare:

{duel_text}

---

Which thesis is stronger overall for the acquirer? Judge objectively, on the strategic case alone, by the following criteria:
- Strategic fit with the acquirer's existing business
- Synergy potential (revenue and cost synergies)
- Feasibility and execution risk
- Value creation potential

You must pick exactly one winner; ties are not allowed.

Respond in the following JSON format:

{
  "comparison_reasoning": "2-3 sentences explaining the key differentiators and your decision.",
  "winner": "A" or "B"
}
\end{lstlisting}

\subsection{US customer targeting (Marketing)}

\begin{lstlisting}[style=prompt,caption={Marketing few-shot hypothesis generation (system).},label={lst:personas-fewshot-gen}]
You are a market research expert who generates hypotheses about users' interest in products and services.
\end{lstlisting}

\begin{lstlisting}[style=prompt,caption={Marketing few-shot hypothesis generation (user).},label={lst:personas-fewshot-gen-user}]
## Target product/service:
{target_product}

---

## Existing example hypotheses (for inspiration only):

{existing_theses}

---

## Persona to evaluate:
{persona_profile}

---

Generate a compelling hypothesis explaining why this persona would be interested in this product.

Focus on the following:
- The person's specific needs or pain points that the product can address
- How the product fits into that person's lifestyle and daily life
- The emotional and practical benefits for this specific persona
- The specific features or attributes that resonate with that person

Refer specifically to the actual persona, citing their background, interests, and situation.

Make the strongest plausible case for why this persona is worth targeting; conclude they are a poor target only if the fit is genuinely poor.

You can use the existing hypotheses as inspiration:
- **Mutate**: Take the reasoning angle of ONE existing hypothesis and adapt it to this persona
- **Crossover**: Combine elements of TWO OR MORE hypotheses to create new logic
- **Explore**: Propose a completely different motivation or use case not covered by any hypothesis

Also report which operation you used, and which of the "Persona N" numbers above you based it on (source_examples). For Explore, set source_examples to an empty list.

Do not copy phrases verbatim from the existing hypotheses.

Respond in the following JSON format:

{
  "operation": "mutate" | "crossover" | "explore",
  "source_examples": [The persona numbers you based it on. E.g., for mutate [2], for crossover [1, 3], for explore []],
  "evolution_note": "One sentence explaining how this hypothesis was influenced by the existing example hypotheses (which part of which 'Persona N' you incorporated, etc.). This is the only field where you may reference the examples. Empty string for explore.",
  "hypothesis": "2-6 sentences explaining why this persona would be interested in this product (or, only if the fit is genuinely poor, why they would not). Write it as a standalone text, as if you had written it without the example hypotheses at hand: do not reference, number, or hint at the example hypotheses, and do not explain in the body which operation you used or how it was derived (those references go only in evolution_note)."
}
\end{lstlisting}

\begin{lstlisting}[style=prompt,caption={Marketing pointwise judge (system).},label={lst:personas-pointwise-judge}]
You are a critical market research analyst scoring persona-targeting recommendations.
\end{lstlisting}

\begin{lstlisting}[style=prompt,caption={Marketing pointwise judge (user).},label={lst:personas-pointwise-judge-user}]
Evaluate the following persona-targeting recommendation.

## Target product/service:
{target_product}

## Persona profile:
{persona_profile}

## Hypothesis (why this persona would or would not be interested):
{hypothesis}

---

Score how compelling and plausible the case for targeting this persona is, considering:
- Relevance: Does the product actually address this persona's needs?
- Specificity: Is the case specific to this persona, or is it generic?
- Plausibility: Is it believable that this persona would actually buy/use the product?
- Depth of insight: Does it demonstrate a genuine understanding of the persona?
- Value alignment: Does the product align with this persona's values, lifestyle, and means?

A hypothesis that argues the persona would NOT be interested, concludes the fit is poor, or offers only a hollow/implausible rationale is a weak recommendation and must score low, even if its reasoning is accurate. Otherwise evaluate strictly and honestly; not every recommendation is good.

Provide your evaluation in the following JSON format:

{
  "reasoning": "A concise one-sentence summary of the evaluation",
  "score": <integer from 0 to 10>
}

Scoring guide (strength of the recommendation to target this persona):
- 0-2: No real case (argues against, no fit, generic reasoning, or not believable)
- 3-4: Weak case (weak fit, serious concerns)
- 5-6: Moderate case (some relevance, but lacking specificity or depth)
- 7-8: Strong case (clear fit, manageable concerns)
- 9-10: Excellent case (compelling, specific, and insightful connection)
\end{lstlisting}

\begin{lstlisting}[style=prompt,caption={Marketing novelty check (system).},label={lst:personas-novelty}]
You are a senior market research expert who assesses the novelty of persona-product hypotheses.
\end{lstlisting}

\begin{lstlisting}[style=prompt,caption={Marketing novelty check (user).},label={lst:personas-novelty-user}]
## Target product/service:
{target_product}

---

## Existing hypotheses:

{existing_text}

---

## New hypothesis to assess:

{new_text}

---

Assess whether the NEW hypothesis above represents a meaningfully different hypothesis compared to the existing ones:
- Does the new hypothesis explore a different motivation or use case for the product?
- Focus on the novelty of the hypothesis, not the difference in users.
- Evaluate strictly: trivial variations are not novel.

Respond in the following JSON format:

{
  "novelty_reasoning": "1-2 sentences explaining whether the new hypothesis is novel compared to the existing ones.",
  "is_novel": true or false
}
\end{lstlisting}

\begin{lstlisting}[style=prompt,caption={Marketing blind pairwise duel (system).},label={lst:personas-duel}]
You are a senior market research expert who compares persona-product hypotheses.
\end{lstlisting}

\begin{lstlisting}[style=prompt,caption={Marketing blind pairwise duel (user).},label={lst:personas-duel-user}]
## Target product/service:
{target_product}

---

## Hypotheses to compare:

{duel_text}

---

Which hypothesis is better overall? Judge objectively, on content quality alone, by the following criteria:
- Relevance: How well does the product fit this persona's needs?
- Specificity: Is the hypothesis specific and insightful?
- Plausibility: Would this persona realistically use the product?
- Value alignment: Does the product align with this persona's values, lifestyle, and means?

You must pick exactly one winner; ties are not allowed.

Respond in the following JSON format:

{
  "comparison_reasoning": "2-3 sentences explaining the key differentiators and your decision.",
  "winner": "A" or "B"
}
\end{lstlisting}

\subsection{Paper-abstract exploration}

\begin{lstlisting}[style=prompt,caption={ML Papers few-shot memo generation (system).},label={lst:abs-fewshot-gen}]
You are a research advisor who writes memos on whether papers are worth reading.
\end{lstlisting}

\begin{lstlisting}[style=prompt,caption={ML Papers few-shot memo generation (user).},label={lst:abs-fewshot-gen-user}]
## Researcher Profile:
{researcher_profile}

---

## Existing Example Memos (for inspiration only):

{existing_recommendations}

---

## New Paper:
Title: {title}

Abstract:
{abstract}

---

Write a concise research memo for this researcher.

The abstract is already a summary, so DO NOT re-summarize the paper. Instead, focus on:
- Why this paper could be interesting for THIS researcher (your "hypothesis" of relevance)
- One concrete next step (experiment, ablation, or extension) if they read it

Use the existing memos as evolutionary building blocks:
- Mutate: adapt a useful reasoning pattern from ONE example to the new paper and this researcher
- Crossover: combine complementary angles from TWO OR MORE examples into a stronger memo
- Explore: propose a genuinely different angle not covered by any example

Also REPORT which operator you used and the example number(s) you built on, using the
"Example N" numbers shown above (leave the source list empty for Explore).

IMPORTANT:
- Do not copy phrases verbatim from the examples.
- Be grounded in the abstract; do not invent unrelated claims.
- Keep it short: 2-6 sentences (one short paragraph).
- Write in plain text; do NOT include LaTeX or math notation.

Format your response as JSON:
{
  "operation": "mutate" | "crossover" | "explore",
  "source_examples": [example numbers you built on, e.g. [2] for mutate or [1, 3] for crossover; [] for explore],
  "evolution_note": "ONE sentence on how the example memo(s) shaped this --- e.g. which angle came from which Example N. This is the ONLY field that may reference the examples; leave empty for explore.",
  "research_notes": "2-6 sentences: start with your relevance hypothesis for this researcher (no paper summary), and end with one concrete next step (experiment, ablation, or extension) grounded in the abstract. Write the memo as STANDALONE: do NOT reference, number, or allude to the example memos, and do NOT narrate the operator or how it was derived (put any such mention only in evolution_note)."
}
\end{lstlisting}

\begin{lstlisting}[style=prompt,caption={ML Papers pointwise judge (system).},label={lst:abs-pointwise-judge}]
You are a critical research advisor who evaluates paper recommendations.
\end{lstlisting}

\begin{lstlisting}[style=prompt,caption={ML Papers pointwise judge (user).},label={lst:abs-pointwise-judge-user}]
## Researcher Profile:
{researcher_profile}

---

## Paper Abstract:
{paper_abstract}

---

## Research Notes:
{research_memo}

---

Evaluate this paper recommendation for the researcher.

Assess how valuable this paper + research memo is for this researcher, considering:
- Relevance to the researcher's stated interests
- Technical quality and novelty
- Quality of insights in the research memo
- Potential for inspiring new research directions

Be rigorous and honest. Most recommendations are NOT outstanding. It might be written as though there are deep connections with the researcher's interests, but don't be fooled easily.

Calibration (important):
- Use the full 0-10 scale. Avoid score inflation.
- If the memo is generic (could fit many papers) or offers vague directions, score it lower even if it "sounds reasonable".
- Prefer concrete, falsifiable, technically grounded next steps over broad brainstorming.

Provide your evaluation in this EXACT JSON format:

{
  "reasoning": "2-3 sentences summarizing your assessment",
  "score": <integer from 0-10>
}

Scoring guide:
- 0-2: Poor recommendation (irrelevant/low-signal; likely a waste of time)
- 3-4: Weak recommendation (thin connection; generic memo; few concrete next steps)
- 5-6: OK / mixed (some relevance, but limited novelty or actionable direction; unsurprising)
- 7: Good (clear relevance + shows specific, plausible research directions)
- 8: Very good (strong relevance + multiple specific directions; minimal hand-waving)
- 9-10: Exceptional (highly relevant, genuinely novel insight, very interesting)
\end{lstlisting}

\begin{lstlisting}[style=prompt,caption={ML Papers novelty check (system).},label={lst:abs-novelty}]
You are a senior research advisor who assesses the novelty of paper recommendations.
\end{lstlisting}

\begin{lstlisting}[style=prompt,caption={ML Papers novelty check (user).},label={lst:abs-novelty-user}]
## Researcher Profile:
{researcher_profile}

---

## Existing Recommendations:

{existing_text}

---

## New Recommendation to Assess:

{new_text}

---

Assess whether the NEW recommendation above offers a meaningfully different reading angle or memo perspective compared to the existing ones.

Consider:
- Does the new memo explore a different relevance hypothesis or research direction for this researcher?
- Focus on the angle of the memo, not just the topic of the paper.
- Evaluate strictly: superficial differences in paper topic with the same reading angle are NOT novel. Only a genuinely different memo framing or research direction counts.

Respond in the following JSON format:

{
  "novelty_reasoning": "1-2 sentences explaining whether the new recommendation is a genuinely novel reading angle or memo framing compared to the existing ones.",
  "is_novel": true or false
}
\end{lstlisting}

\begin{lstlisting}[style=prompt,caption={ML Papers blind pairwise duel (system).},label={lst:abs-duel}]
You are a senior research advisor who compares paper recommendations and decides which is more valuable.
\end{lstlisting}

\begin{lstlisting}[style=prompt,caption={ML Papers blind pairwise duel (user).},label={lst:abs-duel-user}]
## Researcher Profile:
{researcher_profile}

---

## Recommendations to Compare:

{duel_text}

---

Which recommendation is more valuable for this researcher? Judge objectively, on memo content alone, by the following criteria:
- Relevance to the researcher's stated interests
- Specificity and depth of the reading hypothesis
- Quality and concreteness of the proposed research directions
- Potential to inspire new work

You must pick exactly one winner; ties are not allowed.

Respond in the following JSON format:

{
  "comparison_reasoning": "2-3 sentences explaining the key differentiators and your decision.",
  "winner": "A" or "B"
}
\end{lstlisting}

\subsection{Goodreads-Children (grounding)}
Goodreads has no LLM quality judge (Quality is the reader's held-out star rating), so its prompt set is smaller: memo generation and the novelty check. For a non-novel book, replacement compares its human rating directly with that of its nearest neighbor: a strictly higher rating replaces the neighbor, while a tie or lower rating retains the incumbent. The bandit uses rating/5 for both seed initialization and subsequent updates, while reported Quality remains on the 1--5 scale. Previously tried books, including seeds, are excluded from subsequent selection.

\begin{lstlisting}[style=prompt,caption={Goodreads few-shot memo generation (system).},label={lst:goodreads-fewshot-gen}]
You are a reading advisor who writes memos on whether a reader would enjoy a book.
\end{lstlisting}

\begin{lstlisting}[style=prompt,caption={Goodreads few-shot memo generation (user).},label={lst:goodreads-fewshot-gen-user}]
## Reader Profile:
{profile}

---

## Existing Example Memos (for inspiration only):

{existing_recommendations}

---

## New Candidate Book:
Title: {title}

Description:
{description}

---

Write a concise memo on why this reader might (or might not) enjoy this book.

The description already summarizes the book, so DO NOT re-summarize it. Instead, focus on:
- Why THIS reader in particular might enjoy (or not enjoy) this book, grounded in the specific tastes, themes, and pet peeves visible in their profile.
- What about the book connects to or clashes with what they have rewarded or panned before.

Use the existing memos as evolutionary building blocks:
- Mutate: adapt a useful reasoning angle from ONE example to the new book and this reader
- Crossover: combine complementary angles from TWO OR MORE examples
- Explore: propose a genuinely different angle not covered by any example

Also REPORT which operator you used and the example number(s) you built on, so the
reasoning can be traced. Use the "Example N" numbers shown above; for Explore, leave
the source list empty.

IMPORTANT:
- Do not copy phrases verbatim from the examples.
- Be grounded in the book description and the reader's profile; do not invent plot details not in the description.
- Keep it short: 2-6 sentences (one short paragraph).
- Do not mention or guess a star rating.

Format your response as JSON:
{
  "operation": "mutate" | "crossover" | "explore",
  "source_examples": [example numbers you built on, e.g. [2] for mutate or [1, 3] for crossover; [] for explore],
  "evolution_note": "ONE sentence on how the example memo(s) shaped this --- e.g. which angle came from which Example N. This is the ONLY field that may reference the examples; leave empty for explore.",
  "notes": "2-6 sentences: start directly with your relevance hypothesis grounded in this reader's demonstrated tastes; do not re-summarize the plot. Write it as STANDALONE: do NOT reference, number, or allude to the example memos, and do NOT narrate the operator or how it was derived (put any such mention only in evolution_note)."
}
\end{lstlisting}

\begin{lstlisting}[style=prompt,caption={Goodreads novelty check (system).},label={lst:goodreads-novelty}]
You are a reading advisor who assesses whether a new book recommendation adds genuine variety to a reader's shelf.
\end{lstlisting}

\begin{lstlisting}[style=prompt,caption={Goodreads novelty check (user).},label={lst:goodreads-novelty-user}]
## Reader Profile:
{profile}

---

## Recommendations on the shelf:

{recommendations_text}

---

The NEW recommendation is: **{new_recommendation_label}**

Assess whether the NEW recommendation is meaningfully different from the existing ones --- i.e. whether it adds genuine variety to a diverse shelf for this reader.

Important:
- Judge by the kind of book and the kind of appeal, not by quality.
- Be strict: a superficially different title that offers the same kind of reading experience as an existing one is NOT novel.
- Do not use or guess any ratings/scores; judge by content and appeal only.

Provide your response in this EXACT JSON format:

{
  "novelty_reasoning": "1-2 sentences explaining why the new recommendation is or isn't novel.",
  "is_novel": true or false
}
\end{lstlisting}

\section{Example QD-Rec recommendations}
\label{app:examples}
Two QD-Rec recommendations from M\&A and marketing (proposer \texttt{claude-sonnet-5}, judge \texttt{gpt-5.6-terra}). Each was made by the \texttt{mutate} operator from a single parent in the archive, and we chose them to show how a parent hypothesis shapes the child. For each, we give the chosen item, the parent hypothesis, the child hypothesis, and the hypothesis that Novelty wrote for the same item without exemplars. Texts are verbatim apart from the replacement of non-ASCII characters.

\subsection*{M\&A target discovery: JPMorgan Chase \& Co (acquirer)}
\noindent\emph{Acquirer profile: JPMorgan Chase \& Co. Hypothesis = acquisition thesis (judge 0--10). QD-Rec run with seed 1; the child is evaluation 312 of 1{,}000, made by \texttt{mutate} from one parent. Scores: parent 1/10, child \textbf{5/10}, Novelty 1/10 (seed 2). The child entered the archive and was later replaced by a nearby entry that won a pairwise duel against it.}

\needspace{4\baselineskip}
\begin{lstlisting}[style=prompt,caption={Chosen item.}]
Company: Liberty TripAdvisor Holdings Inc
Location: Englewood, United States of America
Sector / Industry: Technology / Online Services
Activity: Online Services (NEC)
Enterprise Value (2024-08-31, local currency): 2083588988.12
YTD / 2Y total return (%): -49.4117647044217 / -35.7825567494573

Business Description:
Liberty TripAdvisor Holdings, Inc. operates as a family of brands that connect people to experiences. The Company leverages its brands, technology, and capabilities to connect its global audience with partners through rich content, travel guidance, and two-sided marketplaces for experiences, accommodations, restaurants, and other travel categories. The Company operates through three segments: Brand Tripadvisor, Viator, and TheFork. The Brand Tripadvisor segment includes Tripadvisor-branded hotels, media and advertising, and Tripadvisor experiences and dining business. The Viator segment provides an online marketplace that allows travelers to research and book tours, activities and attractions in travel destinations through its Viator branded platform, which includes Website, mobile Web, and mobile app. TheFork segment provides information and services for consumers to research and book dining reservations through its dedicated online restaurant reservations platform, TheFork.
\end{lstlisting}
\needspace{4\baselineskip}
\begin{lstlisting}[style=prompt,caption={Parent hypothesis, for Southwest Airlines Co (score 1/10).}]
Southwest Airlines is unusually well-suited to JPMorgan given the bank's existing role as issuer of the Southwest Rapid Rewards co-brand card portfolio, one of the largest travel rewards programs in the country; full ownership would let JPMorgan capture the entire loyalty-program economics--breakage, points sales, and ancillary fee revenue--rather than sharing them through an issuing partnership, mirroring how airline loyalty programs elsewhere have been valued as standalone assets worth more than the carrier itself. At an EV of $15.3B, Southwest is trading below its levels of two ($15.8B) and four years ago ($17.5B) despite a modest 1.3% YTD gain, offering JPMorgan an opportunity to acquire a top-tier domestic airline brand and its 100+ million-member loyalty base at a valuation depressed by sector-wide fuel and margin pressure rather than by any structural decline in Southwest's franchise. Vertical integration would let JPMorgan embed banking, financing, and payments products directly into the travel purchase funnel--installment financing for fares through Chase's balance-sheet lending, bundled Chase Travel booking, and richer transaction data for underwriting and cross-sell into CCB's card and deposit products. At roughly 1.6% of JPMorgan's ~$963B enterprise value, this is a financially digestible bolt-on that diversifies earnings into a consumer-facing, high-frequency transaction business while deepening an already-existing commercial relationship, though it would require JPMorgan to underwrite genuine airline operating risk (fuel, labor, capacity cycles) that sits outside its traditional risk models.
\end{lstlisting}
\needspace{4\baselineskip}
\begin{lstlisting}[style=prompt,caption={QD-Rec child hypothesis (\texttt{mutate}, score 5/10).}]
Liberty TripAdvisor's three marketplaces--Tripadvisor, Viator, and TheFork--represent a global audience of travel researchers and bookers that JPMorgan could plug directly into Chase Travel, its growing proprietary booking platform, and into Chase Offers and Ultimate Rewards redemption, capturing interchange, points-funded bookings, and installment-financing revenue on activities, dining, and accommodations that currently flow to third-party payment rails. At a $2.08B enterprise value--down sharply from $3.39B three years ago and off 49% YTD--the company is trading near a multi-year low, giving JPMorgan an opportunity to acquire a recognizable global travel brand and its two-sided marketplace infrastructure (Viator's tours/activities network, TheFork's restaurant reservation footprint) at a depressed valuation rather than paying peak multiples seen in prior years. Owning these consumer touchpoints would also generate rich transaction and behavioral data to sharpen underwriting and cross-sell into Chase's card and travel-rewards products, while TheFork's European dining density and Viator's experiences marketplace diversify Chase's travel ecosystem beyond flights and hotels. At roughly 0.2% of JPMorgan's ~$963B enterprise value, this is an immaterial-sized, easily digestible bolt-on that extends the bank's travel-commerce strategy into adjacent categories at a cyclically discounted entry point.
\end{lstlisting}
\needspace{4\baselineskip}
\begin{lstlisting}[style=prompt,caption={Novelty hypothesis for the same item, written without exemplars (score 1/10).}]
This deal lacks strategic logic: JPMorgan is a diversified financial holding company (banking, markets, asset management) while Liberty TripAdvisor is a consumer travel marketplace business with no meaningful overlap in capabilities, technology, or customer needs that JPMorgan doesn't already access via its Chase Travel platform and existing card-linked travel partnerships. At roughly $2.1B enterprise value versus JPMorgan's ~$963B EV, the deal is immaterial in size and would not move the needle financially, while Liberty TripAdvisor's deteriorating fundamentals (-49% YTD, -36% over two years, EV down from ~$3.4B four years ago) signal a structurally challenged business facing competitive pressure from Google, Booking Holdings, and Expedia in online travel. JPMorgan's core competitive advantages--balance sheet scale, regulatory relationships, payments infrastructure, and wealth management--do not translate into any obvious right-to-win in managing a travel content and booking marketplace, and the segments (Viator, TheFork, Tripadvisor media) would sit awkwardly within a bank holding company's risk and regulatory framework. Any travel-related synergy JPMorgan seeks (e.g., enhancing Chase Sapphire travel benefits or Chase Travel bookings) can be achieved more efficiently through commercial partnerships or licensing rather than full acquisition and integration of a publicly-troubled, sub-scale asset. We recommend against pursuing this acquisition given the poor strategic fit and absence of compelling synergy rationale.
\end{lstlisting}

Novelty wrote its hypothesis for Liberty TripAdvisor without earlier examples. It argued that JPMorgan's existing Chase Travel platform and commercial partnerships made an acquisition unnecessary, so it offered no affirmative reason to buy the company (1/10). QD-Rec drew on an archived hypothesis proposing that JPMorgan acquire Southwest Airlines. The parent proposed integrating financing, payments and rewards into travel purchases, and using transaction data to improve credit decisions and offer related financial products. The judge recognized this commercial rationale but scored the parent 1/10, citing restrictions on bank ownership of an airline and exposure to fuel, labor and other operating risks.

The \texttt{mutate} operator adapted this rationale to the travel marketplaces described in Liberty TripAdvisor's company information. The child proposed connecting bookings for activities, dining and accommodation to Chase Travel, rewards redemption and installment financing. It retained the parent's idea of linking travel purchases to financial services, while changing the target from an airline to a travel platform. The judge rated the child 5/10, although it still raised concerns about competition, acquisition complexity and whether partnerships would suffice. A parent rated only 1/10 thus served as a stepping stone for a more highly rated recommendation about a different company.

In this example, the model generated statements and interpretations absent from the input, including a loyalty base of more than 100 million members. For example, \href{https://investors.southwest.com/news-events/press-releases/detail/1935/southwest-airlines-reports-second-quarter-2026-results-earnings-power-on-full-display}{Southwest's July 22, 2026 announcement} came after the model's knowledge cutoff and reported ``nearly 100 million Members.'' We could not verify the claim of more than 100 million members. The financial snapshot is dated August 2024, but note that \href{https://www.sec.gov/Archives/edgar/data/1526520/000110465925041251/tm2513424d1_ex99-1.htm}{Tripadvisor acquired Liberty TripAdvisor Holdings in April 2025}. In this experiment, we supplied the model with only basic financial data and short business descriptions. Using the recommendations for actual decisions would require more detailed and up-to-date information.

\subsection*{Marketing: Upskilling (SkillForge)}
\noindent\emph{Product: SkillForge learning platform. Hypothesis = targeting hypothesis (judge 0--10). QD-Rec run with seed 1; the child is evaluation 438 of 1{,}000, made by \texttt{mutate} from one parent. Scores: parent 7/10, child \textbf{7/10}, Novelty 2/10 (seed 3). The child did not enter the archive: it lost the pairwise duel against its nearest archive entry.}

\needspace{4\baselineskip}
\begin{lstlisting}[style=prompt,caption={Chosen item.}]
Jesus Montalvo, a 29-year-old nurse who balances a laid-back, sometimes disorganized lifestyle with a fierce curiosity for evidence-based care finds harmony in solitary hikes, guitar strings and cooking family recipes, even as they occasionally forget bill due dates and misplace a stethoscope.

Age / Sex: 29 / Male Marital status: never_married
Education: some_college
Occupation: registered_nurse Location: National City, CA

Career goals & ambitions:
He aspires to earn a Bachelor of Science in Nursing (BSN) and eventually become a Family Nurse Practitioner, focusing on primary care for underserved Latino communities in San Diego County. He also hopes to take on a charge nurse role to gain more leadership experience, and eventually develop a community outreach program that provides preventive health screenings in his hometown of National City. Balancing flexibility with professional growth, he aims to work part-time or on a per-shift basis to preserve time for personal interests.

Professional persona:
Jesus Montalvo, a 29-year-old registered nurse specializing in acute patient assessment, IV therapy and emergency triage, blends evidence-based curiosity with a relaxed, adaptable work style, often juggling charge-nurse responsibilities while championing Spanish-speaking patient advocacy and quietly steering multidisciplinary teams toward practical outcomes.
\end{lstlisting}
\needspace{4\baselineskip}
\begin{lstlisting}[style=prompt,caption={Parent hypothesis, for Sharon English, 46, preschool or kindergarten teacher (score 7/10).}]
Sharon's explicit ambition to move from the classroom into a district-wide leadership role--developing STEM curricula and mentoring other teachers--requires competencies her ECE and STEM pedagogy training doesn't fully provide: project management to roll out a district-wide program, data analysis to evaluate learning outcomes, and polished presentation skills to advocate for inclusive policies to administrators and school boards. SkillForge's Business Track maps directly onto these gaps, offering practical frameworks she could immediately apply while still pursuing her master's degree. Her comfort with structure, spreadsheets, and hands-on STEM tools also makes the Tech Track's approachable Python and data courses appealing, letting her build lightweight tools or dashboards to track her STEM curriculum's impact--something she could point to when pitching herself for a curriculum specialist role. As someone who already devotes personal time to maker-space workshops and mentoring at an after-school STEM club, the platform's weekly 3-4 hour commitment, live expert Q&As, and LinkedIn-postable certificates fit naturally into her disciplined, growth-oriented routine and would reinforce the professional credibility she needs as she transitions from teaching into educational leadership.
\end{lstlisting}
\needspace{4\baselineskip}
\begin{lstlisting}[style=prompt,caption={QD-Rec child hypothesis (\texttt{mutate}, score 7/10).}]
Jesus's stated goal of becoming a charge nurse and eventually launching a preventive health outreach program in National City requires competencies his clinical training never covered--project management to coordinate a screening initiative, basic data analysis to track and report health outcomes, and polished presentation skills to pitch the program to hospital administrators or grant funders. SkillForge's Business Track speaks directly to that gap, while the Tech Track's approachable data courses could let him build simple dashboards to demonstrate the impact of his community screenings, reinforcing his evidence-based, protocol-driven mindset. His per-shift, flexible work schedule and self-described tendency toward disorganization make the platform's structured 3-4 hour weekly cadence, dashboard progress tracking, and cancel-anytime plan especially appealing, since it offers accountability without locking him into a rigid semester-style commitment. A LinkedIn-postable certificate would also lend credibility as he positions himself for charge-nurse responsibilities and eventually a leadership role serving underserved Latino patients. While SkillForge can't replace the BSN or FNP coursework he ultimately needs, it fills a real and immediate void in the program-management and advocacy skills his broader career vision demands, making him a solid niche fit despite the platform's non-clinical focus.
\end{lstlisting}
\needspace{4\baselineskip}
\begin{lstlisting}[style=prompt,caption={Novelty hypothesis for the same item, written without exemplars (score 2/10).}]
Jesus's near-term career goals are almost entirely clinical--finishing his BSN, becoming an FNP, and building a community health outreach program--none of which SkillForge's business, tech, or creative tracks can substitute for, since it offers no accredited nursing or healthcare credentialing. While he might be mildly drawn to the Business Track's project management or data analysis modules to sharpen his charge-nurse leadership and eventually structure a preventive-screening program, this is a peripheral use case rather than a core need, and his already packed per-shift schedule plus stated preference for unstructured downtime (hiking, guitar, cooking) makes committing another 3-4 hours weekly to a subscription-based platform a hard sell. Because the platform's value proposition of 'job-ready skills for white-collar/tech/creative careers' doesn't map onto his actual professional trajectory, and he shows no expressed interest in side-gig income streams like UI/UX or video editing, SkillForge is a poor strategic fit for this persona despite superficial overlap in the 'practical skill-building' ethos.
\end{lstlisting}

Novelty focused on the nurse's clinical career goals and SkillForge's lack of accredited nursing courses. It recognized that project management and data analysis could help with his planned screening program, but treated these as peripheral needs and his shift work as an obstacle to study. It concluded that the platform was a poor fit (2/10). QD-Rec drew on an archived hypothesis about a preschool teacher seeking district-wide curriculum leadership. The parent argued that SkillForge could supply project management, data analysis and presentation skills needed for this transition but not fully covered by her teaching training (7/10).

The \texttt{mutate} operator applied this reasoning to the nurse's ambition to lead a team and launch a community screening program. The child connected project management to coordinating screenings, data analysis to reporting health outcomes, and presentation skills to securing support from hospital administrators or funders. These skills became the main reason to recommend SkillForge as a complement to his clinical education. The child also argued that the platform's flexible format could suit his irregular schedule. The judge found the connection between his leadership goals and the courses credible and rated the child 7/10. Reusing the teacher's hypothesis helped QD-Rec connect the nurse's leadership goals to courses that Novelty had dismissed as peripheral.

\end{document}